\documentclass[prd,showpacs,eqsecnum]{revtex4}
\usepackage{bm}
\usepackage{amssymb}
\usepackage{amsmath}

\begin{document}
\title{Self-forces on extended bodies in electrodynamics}
\author{Abraham I. Harte}
\affiliation{Institute for Gravitational Physics and Geometry,
\\
Center for Gravitational Wave Physics, Department of Physics,
\\
The Pennsylvania State University, University Park, PA 16802}

\date{January 16, 2006}


\begin{abstract}
In this paper, we study the bulk motion of a classical extended
charge in flat spacetime. A formalism developed by W. G. Dixon is
used to determine how the details of such a particle's internal
structure influence its equations of motion. We place essentially no
restrictions (other than boundedness) on the shape of the charge,
and allow for inhomogeneity, internal currents, elasticity, and
spin. Even if the angular momentum remains small, many such systems
are found to be affected by large self-interaction effects beyond
the standard Lorentz-Dirac force. These are particularly significant
if the particle's charge density fails to be much greater than its
3-current density (or vice versa) in the center-of-mass frame.
Additional terms also arise in the equations of motion if the dipole
moment is too large, and when the `center-of-electromagnetic mass'
is far from the `center-of-bare mass' (roughly speaking). These
conditions are often quite restrictive. General equations of motion
were also derived under the assumption that the particle can only
interact with the radiative component of its self-field. These are
much simpler than the equations derived using the full retarded
self-field; as are the conditions required to recover the
Lorentz-Dirac equation.

\end{abstract}

\maketitle

\vskip 2pc

\section{Introduction}

Originally motivated by the discovery of the electron, the behavior
of small (classical) electric charges has been studied in various
contexts for over a century. The first results were obtained by
Abraham \cite{Abraham} for a non-relativistic non-spinning rigid
sphere. This calculation was later repeated by Schott \cite{Schott}
and Lorentz \cite{Lorentz} within special relativity, and by Crowley
and Nodvik \cite{NodvikOrig} in general (background) spacetimes.
Similar results have since been obtained by a number of authors for more general charge distributions \cite{Schott1, Nodvik, Yaghjian, Jackson, Barut, Crisp, OriExtend} in flat spacetime. These results have very recently been extended to curved backgrounds as well \cite{PoissonExtend}. Still, these derivations did not allow for significant elasticity, charge-current coupling,
and/or rotation.

Detailed reviews of different aspects of the self-force problem (in
electromagnetism as well as scalar field theory and general
relativity) have been given by Poisson \cite{SFReview1}, Havas
\cite{SFReview2}, and Spohn \cite{SFReview3}. To summarize, though,
a common theme throughout all of these works has been that the
equations of motion describing sufficiently small particles were
found to be independent of the details of their internal structure
to a considerable degree of precision. Only a few parameters such as
the rest mass and total charge entered into these equations.
Ignoring the effects of spin and internal currents, the apparently
universal correction to the Lorentz force law is given by the
well-known Lorentz-Dirac self-force. If a particle with charge $q$
has a center-of-mass position $z^{a}(\bar{s})$ (with $\bar{s}$ being
a proper time), then this self-force is equal to (using a metric
with signature $-2$, and units in which $c=1$)
\begin{equation}
\frac{2}{3} q^{2} \Big( \dddot{z}^{a}(\bar{s}) +
\dot{z}^{a}(\bar{s}) \ddot{z}^{b}(\bar{s}) \ddot{z}_{b}(\bar{s})
\Big) ~. \label{LD}
\end{equation}

Although this is only an approximation to the full self-force for
any realistic (extended) charge, it is natural to introduce a new
concept into the theory for which it is exact -- that of a point
particle. The immediate problem with this is of course that the
self-field diverges at the location of any point-like source, which
would appear to imply that its self-force and self-energy are not
well-defined. Dirac \cite{Dirac} removed this problem by noting that
self-forces acting on a structureless particle should only arise as
a reaction to emitted radiation. Letting $F^{ab}_{\mathrm{self}(+)}$
denote the advanced self-field and $F^{ab}_{\mathrm{self}(-)}$ the
retarded one, we can define `radiative' and `singular' portions of
$F^{ab}_{\mathrm{self}(-)}=F^{ab}_{\mathrm{self}(S)}+F^{ab}_{\mathrm{self}(R)}$:
\begin{eqnarray}
F^{ab}_{\mathrm{self}(S)} &=& \frac{1}{2} \left(
F^{ab}_{\mathrm{self}(-)} + F^{ab}_{\mathrm{self}(+)} \right) ~,
\\
F^{ab}_{\mathrm{self}(R)} &=& \frac{1}{2} \left(
F^{ab}_{\mathrm{self}(-)} - F^{ab}_{\mathrm{self}(+)} \right) ~.
\label{radfielddefine}
\end{eqnarray}
As the names imply, the singular field contains the entire divergent
part of the retarded field. It is also derived from a time-symmetric
Green function, so it would not be expected to contain any
radiation. The self-force on a point particle should therefore be
determined entirely by the radiative self-field. Combining it with
the Lorentz force law immediately recovers (\ref{LD}) \cite{Dirac}.

This prescription is much simpler than any direct derivation of the
equations of motion for a finite charge, and for this reason, it has
been generalized to work in curved spacetime, as well as for scalar
and gravitational self-forces \cite{detwhiting,SFReview1} (although
this is not the only way of `renormalizing' point particle
self-fields \cite{QuinnWald,Quinn}). As before, the relevance of
these extensions to realistic extended bodies has also been
established in certain special cases. In linearized gravity, for
example, a small nearly-Schwarzchild black hole has been found to
obey the same equations of motion as a point particle
\cite{MST,SFReview1}. More generally, it has been shown that a
nonspinning body's internal structure is irrelevant to very high
order within Post-Newtonian theory \cite{Damour}. Still, there
remain questions of exactly how universal these results are. Will a
spherical neutron star in near-equilibrium fall into a supermassive
black hole in the same way as another one that is spinning rapidly
and experiencing internal oscillations \cite{Instabilities}; or one
having a mountainous (solid) surface \cite{NeutronMountains}? Such
systems could be important sources for the upcoming Laser
Interferometer Space Antenna (LISA).

Rather than addressing such questions directly, we have chosen to
study the motions of charged bodies in flat spacetime (in part) as a
model problem. The methods used here were specifically chosen so
that very few conceptual changes would be required to consider
particles (charged or not) moving in a fully dynamic spacetime. This
has led to some additional complexity not strictly necessary to
solve the problem at hand, although the majority of the complicating
issues have been placed in the the appendix.

Our derivation is based on W. G. Dixon's multipole formalism
\cite{Dix67, Dix70a, Dix70b, Dix74, Dix79}. This gives a relatively
simple, unified, and rigorous way of understanding the motions of
arbitrarily structured bodies in both electrodynamics and general
relativity. Despite the fact that Dixon's theory decomposes the
source functions into multipole moments, we never ignore any of
them. By using generating functions, the entire infinite set is
retained throughout all of our calculations. This only appears to be
possible in this formalism. The laws of motion that are used are
therefore exact (despite being ordinary differential equations). All
of the approximations that we make are only used to compute the
self-field.

Taking a hybrid point of view where an extended body can only
interact with its radiative self-field, it is found that the
Lorentz-Dirac equation follows for a wide range of nonspinning
charges with small dipole moments. If the body instead interacts
with its full retarded self-field, this result no longer holds. In
this case, the equations of motion are drastically different if the
charge and (3-) current densities have remotely similar magnitudes
in the center-of-mass frame. Although it is impractical to define
exactly what this means at this point, it will be shown that one of
these quantities must be at least `second order' compared to the
other in order for these extra terms to vanish. This is because
these cases usually allow the self-fields to do significant amounts
of internal work (e.g. Ohmic heating). Even when the charge-current
coupling in negligible, there are still extra complications in the
retarded case. The conditions required to generically exclude these
and other complicating effects are derived, and turn out to be
surprisingly restrictive.

Sec. \ref{ProbMot} reviews the various steps involved in calculating
the motion of matter interacting with an electromagnetic field. Sec.
\ref{LawsMot} then summarizes the appropriate definitions of the
center-of-mass and its laws of motion as obtained from Dixon's
formalism. It also decomposes the 4-current in a particular way that
happens to be convenient in this framework. Although this reduction
is not strictly required for the current problem, it is adopted
throughout on the grounds that it would be essential in curved
spacetime. It is derived in detail in the appendix, which also
contains an in-depth review of of Dixon's ideas.

With these basic ideas in place, Sec. \ref{FieldSect} goes on to
derive expansions for the advanced and retarded self-fields of an
arbitrarily-structured charge using a slow motion approximation.
Sec. \ref{ForceSect} then combines these results with those from
Sec. \ref{LawsMot} to find general expressions for the self-force
and self-torque. Finally, Sec. \ref{SpecCaseSect} examines the
equations of motion for certain simpler classes of charges, and
derives some conditions under which the Lorentz-Dirac equation is
applicable.

We use units in which $c = 1$ throughout. In order to facilitate a
more direct comparison to Dixon's papers, the metric is chosen to
have signature $-2$ (although the rest of the notation used here
frequently differs from Dixon's). We also assume that the spacetime
is flat, and adopt Minkowski coordinates for simplicity. Latin
indices refer to these coordinates, while Greek ones are triad
labels running from $1$ to $3$.

\section{The Problem of Motion}
\label{ProbMot}

In studying the dynamics of any system in a classical field theory,
one has to specify `laws of motion' for both the field and matter
variables. In our case, the only field is of course the
electromagnetic one, $F^{ab}=F^{[ab]}$. As usual, this is governed
by Maxwell's equations
\begin{eqnarray}
\partial_{b}F^{ab} = J^{a} ~,
\label{maxwell}
\\
\partial^{[a} F^{bc]}=0 ~.
\end{eqnarray}

We assume that the matter in our problem is completely described by
its stress-energy tensor $T^{ab}$ and 4-current vector $J^{a}$.
Taking the divergence of (\ref{maxwell}) immediately gives our first
constraint on these quantities:
\begin{equation}
\partial_{a}J^{a}=0 ~.
\label{chargecons}
\end{equation}
This equation acts as one of the laws of motion for the matter
fields. The other is derived from the requirement that
$\partial_{a}\left( T^{ab} + T^{ab}_{\mathrm{em}} \right) =0 $,
where $T^{ab}_{\mathrm{em}}$ is the stress-energy tensor of the
electromagnetic field. Combining the standard form of
$T^{ab}_{\mathrm{em}}$ \cite{Jackson} with Maxwell's equations then
shows that the matter moves according to
\begin{equation}
\partial_{b}T^{ab}=-F^{ab}J_{b} ~.
\label{stresscons}
\end{equation}
(\ref{maxwell})-(\ref{stresscons}) are essentially the entire
content of classical continuum mechanics in flat spacetime. Of
course, different types of matter do not all move in the same way,
so these equations by themselves are not sufficient to determine
$T^{ab}$ and $J^{a}$ for all time (even if the $F^{ab}$ were given).
One also needs to specify something analogous to a (generalized)
equation of state, which can take on a rather unwieldy form.

This sort of procedure is the standard one in continuum mechanics.
Unfortunately, the resulting nonlinear partial differential
equations are notoriously difficult to solve. Such a detailed
description of the system should not really be required, however,
for problems where we are only interested in the body's bulk motion.
In these cases, a representative world line could be defined inside
the (convex hull of the) spacelike-compact support of $T^{ab}$.
Given that this worldline can always be parametrized by a single
quantity, its tangent vector might be expected to satisfy an
ordinary differential equation -- at least when using certain
approximations. Solving such an equation would clearly be much more
straightforward than the partial differential equations that we
started with.

This sort of simplification is one of the main motivations behind
the many (source) multipole formalisms in the literature
\cite{Jackson,ThorneMult}. In these, one first fixes some particular
reference frame which has, among other properties, a preferred time
parameter $s$. The quantity being expanded -- say $J^{a}(x)$ -- can
then be written in terms of an infinite set of tensors depending
only on $s$: $Q^{a}(s)$, $Q^{ab}(s),\ldots$ The reverse is also
true. Given $J^{a}$, there is a well-defined way to compute any
moment. The set $\{Q^{\ldots} \}$ is therefore completely equivalent
to $J^{a}$. This implies that the conservation equation
(\ref{chargecons}) may be used to find restrictions on the
individual moments.

Such restrictions depend on the precise definitions that are being
adopted, but can usually be divided into two general classes. The
first of these consists of purely algebraic equations imposed at a
fixed value of $s$. We call these the constraints. There are also a
number of evolution equations which usually take the form of
ordinary differential equations. Multipole expansions can therefore
be used to convert (\ref{chargecons}) and (\ref{stresscons}) into a
number of algebraic and ordinary differential equations (without any
approximation).

This does not actually simplify things as much as it at first might
appear. The reason is that almost all definitions for the source
multipoles will lead to an infinite number of (coupled!) evolution
equations. This is often dealt with in practice by assuming that all
moments above a particular order are irrelevant, which leaves one
with only a finite number of evolution equations. There are,
however, interesting questions that require knowing the higher
moments. $J^{a}$ (or $T^{ab}$) for example, cannot be reconstructed
without them. This means that the self-field cannot be calculated in
the near zone from only the first few moments. Although the
self-force and self-torque can be found in certain cases by
examining energy and momentum fluxes in the far zone
\cite{Dirac,DewittBrehme,SFReview1,MST} (which can be adequately
approximated using only a finite number of moments), this is
considerably less accurate than integrating the force density
throughout the charge's interior.

For these reasons and others, it is desirable to define a set of
multipoles that do not require any cutoff. Remarkably, such a set
exists \cite{Dix67}, and is essentially unique \cite{Dix74}. Without
any approximation, moments for both $J^{a}$ and $T^{ab}$ can be
defined which satisfy a \textit{finite} number of evolution
equations. There remain (uncoupled) constraint equations for each
moment, although these are easily solved. We adopt this formalism
due to Dixon for the remainder of this paper. Its net effect is to
allow us to relate the motion to the fields in a more rigorous way
than is usually done (short of directly solving (\ref{chargecons})
and (\ref{stresscons})). It does not, however, have anything to say
about the fields themselves. We therefore obtain the self-field in a
standard way, and then use Dixon's equations to find how the matter
moves in response to it.

\section{Laws of Motion}
\label{LawsMot}

As noted, we use Dixon's method \cite{Dix67, Dix70a, Dix70b, Dix74,
Dix79} to decompose $J^{a}$ and $T^{ab}$ into multipole moments.
Each of these sets is designed to describe as simply as possible all
possible forms of $J^{a}$ and $T^{ab}$ satisfying their respective
conservation equations. We first assume that these matter fields are
at least piecewise continuous, and have (identical) supports $W$.
Any intersection of a spacelike hypersurface with this worldtube is
assumed to be compact.

Now choose a timelike worldline $Z \subset W$, and a timelike unit
vector field $n^{a}(s)$ defined on $Z$. It is assumed that this is
always possible in any physically interesting system. $Z$ is then
parameterized by the coordinate function $z^{a}(s)$, and the tangent
vector to it is denoted by $v^{a}(s) := \mathrm{d} z^{a} /
\mathrm{d} s =: \dot{z}^{a}(s)$. $v^{a}$ need not be equal to
$n^{a}$, although it will be convenient to normalize $s$ such that
$n^{a}v_{a}=1$ (so $v^{a}v_{a} \neq 1$ in general). $v^{a}$ is
called the kinematical velocity, while $n^{a}$ is the dynamical
velocity. The set $\{ Z,n^{a} \}$ then defines a reference frame for
the definition of the multipole moments. At this point, it should be
thought of as arbitrary, although physical conditions will later be
given that pick out a unique `center-of-mass' frame.

A collection of spacelike hyperplanes $\{ \Sigma \}$ can easily be
constructed from $Z$ and $n^{a}$. Each $\Sigma(s)$ is to pass
through $z^{a}(s)$, and be (everywhere) orthogonal to $n^{a}(s)$.
Assume that any point in $W$ is contained in exactly one of these
planes. Unless $n^{a}$ is a constant, it is clear that this property
cannot be true throughout the entire spacetime. For $x \in \Sigma(s)
\cap W$, we must therefore have that $\max|\dot{n}^{a}(s)
\big(x-z(s)\big)_{a}| < 1$ (among other conditions), which gives a
weak restriction on the body's maximum size. It is not really
important, though, as any reasonable type of matter would be ripped
apart long before this condition was violated.

The main results that we need from Dixon's theory at this point are
his definitions of the linear and angular momenta. These disagree
with the usual ones when either $J^{a}$ or $F^{ab}$ are nonzero,
although it is still convenient to label them by $p^{a}(s)$ and
$S^{ab}(s)$ respectively. Unless otherwise noted, the words `linear
and angular momenta' will always refer to the quantities
\cite{Dix67, Dix70a}
\begin{eqnarray}
p^{a}(s) &=& \int_{\Sigma(s)} \! \mathrm{d}\Sigma_{b} \, \left[
T^{ab} + J^{b} r_{c} \int_{0}^{1} \mathrm{d}u \; F^{ac}\big( z(s) +
ur \big) \right] ~,
\label{pdefine}
\\
S^{ab}(s) &=& 2 \int_{\Sigma(s)} \! \mathrm{d}\Sigma_{c} \, \left[
r^{[a} T^{b]c} + J^{c} r_{d} \int_{0}^{1} \mathrm{d}u \; u r^{[a}
F^{b]d} \big( z(s) + ur \big) \right]  ~, \label{Sdefine}
\end{eqnarray}
where $r^{a} := x^{a}-z^{a}(s)$, and $u$ is just a dummy parameter
used to integrate along the line segment connecting $z^{a}(s)$ to
$x^{a}=z^{a}(s)+r^{a}$.

Detailed motivations for these definitions can be found in
\cite{Dix67,Dix70a,Dix74}, as well as the appendix. In short,
though, it can be shown that the given quantities are uniquely
determined by demanding that stress-energy conservation directly
affect only the first two moments of $T^{ab}$ (once the concept of a
moment has been defined in a reasonable way). If $p^{a}$ and
$S^{ab}$ are known in some time interval and satisfy the appropriate
evolution equations, the class of all stress-energy tensors with
these moments can be constructed without having to solve any
differential equations. Each of these will exactly satisfy
(\ref{stresscons}).

This property implies that evolution equations for the quadrupole
and higher moments of the stress-energy tensor are nearly
unconstrained. They can be thought of as the `equation of state' of
the material under consideration. This type of independence of the
higher moments from the conservation laws also occurs in Newtonian
theory \cite{Dix79}, and there are considerable advantages in
preserving as much of that structure as possible in the relativistic
regime. In particular, there is no need to discard multipole moments
above a certain order. The choices (\ref{pdefine}) and
(\ref{Sdefine}) allow us to retain many of the conveniences of a
multipole formalism without its classic limitations.

The same definitions for the momenta can also be motivated by
considering charged particles in curved spacetime \cite{Dix70a,
Dix79}. There, one can study the conserved quantities associated
with Killing vectors in appropriate spacetimes (where both the
metric and electromagnetic field are assumed to share the same
symmetries). Fixing $Z$ and $n^{a}(s)$ allows each such quantity to
be written as a linear combination of vector and antisymmetric rank
2 tensor fields on $Z$. Crucially, the definitions of these
quantities do not depend on the Killing vector under consideration,
so we can suppose that they are meaningful even in the absence of
any symmetries. If the metric is taken to be flat, these objects
reduce to (\ref{pdefine}) and (\ref{Sdefine}) \cite{Dix70a}. They
are therefore the natural limits of what would generally be referred
to as the linear and angular momenta of symmetric spacetimes.

It is now convenient to define a coordinate system on $W$ that is
more closely adapted to the system than the Minkowski coordinates
used so far. First choose an orthonormal tetrad $\{ n^{a}(s),
e^{a}_{\alpha}(s) \}$ ($\alpha =1,2,3$) along $Z$. Requiring that
each of these vectors remain orthonormal to the others implies that
\begin{equation}
\dot{e}^{a}_{\alpha} = -n^{a} \dot{n}_{b} e_{\alpha}^{b} ~,
\label{tetrad}
\end{equation}
which is essentially Fermi-Walker transport. A spatial (rotation)
term may also be added to this, although our calculations would then
become considerably more tedious. On its own, the choice of tetrad
has no physical significance, so we choose the simplest case.

Since it was assumed that $\{\Sigma\}$ foliates $W$, any point $x
\in \Sigma(s) \cap W$ may now be uniquely written in terms of $s$
and a `triad radius' $r^{\alpha}$
\begin{equation}
x^{a}=e^{a}_{\alpha}(s) r^{\alpha} + z^{a}(s) ~. \label{XtoRS}
\end{equation}
Varying $r^{\alpha}$ with a fixed value of $s$ clearly generates
$\Sigma(s)$. Also, the Jacobian of the coordinate transformation
$(x^{a}) \rightarrow (r^{\alpha},s)$ is equal to the lapse $N$ of
the foliation,
\begin{eqnarray}
N(x) &=& 1- \dot{n}_{a}(s) r^{a} ~, \nonumber
\\
&=& 1- \dot{n}_{a}(s) e^{a}_{\alpha}(s) r^{\alpha} ~. \label{lapse}
\end{eqnarray}

We will be extensively transforming back and forth between these two
coordinate systems, so it is convenient to abuse the notation
somewhat by writing $f\left(r^{\alpha},s\right)=f\left(x^{a}\right)$
for any function $f$. The intended dependencies should always be
clear from the context. It is also useful to denote quantities such
as $\dot{n}_{a} e^{a}_{\alpha}$ by $\dot{n}_{\alpha}$. Note that in
this notation, $\ddot{n}_{\alpha} = \ddot{n}_{a} e^{a}_{\alpha} \neq
\mathrm{d} \dot{n}_{\alpha} / \mathrm{d} s$.

Using these conventions, it is natural to split $J^{a}$ into the
charge and 3-current densities seen by an observer at $r=0$ (the
`center-of-mass observer')
\begin{equation}
J^{a} = \rho n^{a} + e^{a}_{\alpha} j^{\alpha} ~. \label{Jtetrad}
\end{equation}

It is then shown in the appendix that there exist `potentials'
$\varphi$ and $H^{\alpha}$ which generate $J^{a}$ through the
equations
\begin{eqnarray}
\rho &=& \partial_{\alpha} \left( r^{\alpha} \varphi \right) ~,
\label{chargedensity}
\\
j^{\alpha} &=& N^{-1} \Big[ H^{\alpha} + v^{\beta}
\partial_{\beta} \left( r^{\alpha} \varphi \right) - r^{\alpha} \dot{\varphi} \Big] ~.
\label{threecurrent}
\end{eqnarray}

Denoting the total charge by $q$ and letting $|r|^{2} := -r^{\alpha}
r_{\alpha} \geq 0$, $\varphi\left(r^{\alpha},s\right)$ was found to
be continuous, and equal to $q/4\pi |r|^{3}$ outside $W$. Similarly,
$H^{\alpha}\left(r^{\beta},s\right)$ is given by (\ref{HBoundary})
outside $W$, and is piecewise continuous in $r^{\beta}$.
$H^{\alpha}$ also satisfies $\partial_{\alpha} H^{\alpha} =0$. These
properties guarantee that $J^{a}$ has support $W$, and is everywhere
continuous. A direct calculation also shows that
\begin{eqnarray}
\partial_{a} J^{a} &=& N^{-1} n_{b} \left( \frac{\partial}{\partial s} - v^{\beta} \partial_{\beta} \right) J^{b} + \partial_{\beta} j^{\beta} ~,
\\
&=& 0 ~,
\end{eqnarray}
as required by (\ref{chargecons}).

Any physically reasonable current vector can now be constructed by
choosing potentials satisfying these rules. A physical
interpretation of one's choice is then given by substitution into
(\ref{chargedensity}) and (\ref{threecurrent}). For example, a
uniform spherical charge distribution with time-varying radius
$D(s)$ is described by (assuming $n^{a}=v^{a}$)
\begin{eqnarray}
\varphi(r,s) &=& \frac{q}{4\pi D^{3}(s)} \left[ \Theta \big(
D(s)-|r| \big) + \left( \frac{D(s)}{|r|} \right)^{3} \Theta \big(
|r|-D(s) \big) \right] ~, \label{phiEx}
\\
H^{\alpha}(r,s) &=& 0 ~, \label{HEx}
\end{eqnarray}
where $\Theta(\cdot)$ is the Heaviside step function.
(\ref{chargedensity}) and (\ref{threecurrent}) then show that the
tetrad components of $J^{a}$ are
\begin{eqnarray}
\rho(r,s) &=& \frac{3q}{4\pi D^{3}} \Theta \big( D-|r| \big) ~,
\\
j^{\alpha}(r,s) &=& \frac{3 q r^{\alpha}}{4\pi N(r,s) D^{3}(s)}
\left( \frac{\dot{D}(s)}{D(s)} \right) \Theta \big( D(s)-|r| \big)
~.
\end{eqnarray}

It is also clearly possible to calculate $\varphi$ and $H^{\alpha}$
from any given $J^{a}$. This requires inverting
(\ref{chargedensity}), which acts as a partial differential equation
for $\varphi$ on each time slice. The solution to this equation
would usually have to be obtained numerically, which is clearly
inconvenient. Largely for this reason, we shall consider $\varphi$
and $H^{\alpha}$ to be the given quantities for the remainder of
this paper. $J^{a}$ can be derived from them using operations no
more complicated than differentiation.

Another reason for this unconventional choice is that $\varphi$ and
$H^{\alpha}$ contain all of the multipole moments of $J^{a}$ in a
natural way. As shown in the appendix, they are closely related to
the Fourier transform of a generating function for these moments. An
arbitrary current moment can be obtained essentially by
differentiating the inverse Fourier transforms of the potentials a
suitable number of times. For example, (\ref{QJ}), (\ref{hatJ}),
(\ref{Cform}), (\ref{rhoA}), and (\ref{Bform}) can be used to show
that the dipole moment has the general form
\begin{equation}
Q^{ab} = \int \! \mathrm{d}^{3} r \left[ 2 n^{[a} e^{b]}_{\beta}
r^{\beta} N \left( \varphi(r,s) - \frac{q}{4 \pi |r|^{3}} \right) +
e^{a}_{\alpha} e^{b}_{\beta} \bar{H}^{\alpha \beta}(r,s) \right] ~,
\label{dipole}
\end{equation}
where $\bar{H}^{\alpha \beta} = \bar{H}^{[\alpha \beta]}$ is defined
by
\begin{equation}
H^{\alpha} = \partial_{\beta} \bar{H}^{\alpha \beta} ~. \label{divH}
\end{equation}
Given (\ref{HBoundary}), we let
\begin{equation}
\bar{H}^{\alpha \beta} = \frac{q v^{[\alpha} r^{\beta]} }{2 \pi
|r|^{3} }
\end{equation}
outside $W$.

For the example given in (\ref{phiEx}) and (\ref{HEx}), the dipole
moment is equal to
\begin{equation}
Q^{ab} = \frac{1}{5} q D^{2}(s) n^{[a} \dot{n}^{b]} ~.
\label{dipoleEx}
\end{equation}
This might have been expected to vanish in spherical symmetry,
although it should be noted that it is only nonzero when sphericity
is being defined in an accelerated reference frame.

These constructions can now be used to simplify the definitions of
$p^{a}$ and $S^{ab}$. Letting $\bar{r}^{\alpha} := u r^{\alpha}$,
the electromagnetic term in (\ref{pdefine}) is equivalent to
\begin{equation}
\int \mathrm{d}^{3} \bar{r} \, e_{b}^{\beta} \bar{r}_{\beta}
F^{ab}(\bar{r},s) \int_{0}^{1} \mathrm{d}u \, u^{-4}
\rho(\bar{r}/u,s)~.
\end{equation}
But (\ref{chargedensity}) shows that
\begin{eqnarray}
\int_{0}^{1} \mathrm{d} u \, u^{-4} \rho(\bar{r}/u,s) &=&
-\int_{0}^{1} \mathrm{d} u \, \frac{\partial}{\partial u} \left[
u^{-3} \left( \varphi(\bar{r}/u,s) - \frac{q}{4\pi |\bar{r}/u|^{3}}
\right) \right] ~,
\\
&=& - \left( \varphi(\bar{r},s) - \frac{q}{4\pi |\bar{r}|^{3} }
\right) ~,
\end{eqnarray}
so (\ref{pdefine}) and (\ref{Sdefine}) can be rewritten as
\begin{eqnarray}
p^{a} &=& \int \mathrm{d}^{3} r \left[ T^{ab} n_{b} - \left( \varphi
- \frac{q}{4\pi |r|^{3}} \right) r_{\gamma} e_{c}^{\gamma} F^{ac}
\right] ~, \label{pdefine2}
\\
S^{ab} &=& 2 \int \mathrm{d}^{3} r \left[ r^{\alpha} e^{[a}_{\alpha}
T^{b]c} n_{c} - \left( \varphi - \frac{q}{4\pi |r|^{3}} \right)
r_{\gamma} e_{c}^{\gamma} r^{\alpha} e_{\alpha}^{[a} F^{b]c} \right]
~. \label{Sdefine2}
\end{eqnarray}

We now need evolution equations for these quantities, which are most
easily obtained by direct differentiation. For any function
$I^{b}(x)$, relating the $x$-coordinates of $( r^{\alpha},
s+\mathrm{d} s )$ to those of $( r^{\alpha}, s )$ shows that
\begin{equation}
\frac{\mathrm{d}}{\mathrm{d} s} \int_{\Sigma(s)} \mathrm{d}
\Sigma_{b} I^{b}(x) = \int \mathrm{d}^{3} r \Big[ \dot{n}_{b} I^{b}
+ \left( v^{c} - n^{c} \dot{n}_{\alpha} r^{\alpha} \right)
\partial_{c} \left( n_{b} I^{b} \right)  \Big] ~.
\end{equation}
If $I^{b}$ vanishes outside of some finite radius, this expression
simplifies to
\begin{equation}
\frac{\mathrm{d}}{\mathrm{d} s} \int_{\Sigma(s)} \mathrm{d}
\Sigma_{b} I^{b}(x) = \int \mathrm{d}^{3} r N \partial_{b} I^{b} ~.
\end{equation}
(\ref{stresscons}), (\ref{pdefine2}), and (\ref{Sdefine2}) can now
be used to show that
\begin{eqnarray}
\dot{p}^{a} &=& - \int \mathrm{d}^{3} r \, \Bigg\{ N F^{ab} J_{b} +
\frac{
\partial }{ \partial s} \left[ \left( \varphi
- \frac{q}{4\pi |r|^{3}} \right) r_{\gamma} e_{c}^{\gamma} F^{ac}
\right] \Bigg\} ~, \label{dotP1}
\end{eqnarray}
and
\begin{eqnarray}
\dot{S}^{ab} &=& - 2 \int \mathrm{d}^{3} r \, \Bigg\{ N r^{\alpha}
e^{[a}_{\alpha} F^{b]c} J_{c} + v^{[a} T^{b]c} n_{c}  +
\frac{\partial}{ \partial s} \left[  \left( \varphi - \frac{q}{4\pi
|r|^{3}} \right) r_{\gamma} e_{c}^{\gamma} r^{\alpha}
e_{\alpha}^{[a} F^{b]c} \right] \Bigg\} ~, \nonumber
\\
&=& 2 p^{[a} v^{b]} - 2 \int \mathrm{d}^{3} r \, \Bigg\{  N
r^{\alpha} e^{[a}_{\alpha} F^{b]c} J_{c} + \frac{\partial}{ \partial
s} \left[ \left( \varphi - \frac{q}{4\pi |r|^{3}} \right) r_{\gamma}
e_{c}^{\gamma} r^{\alpha} e_{\alpha}^{[a} F^{b]c} \right] \nonumber
\\
&& ~ +  \left( \varphi - \frac{q}{4\pi |r|^{3}} \right) r_{\gamma}
e_{c}^{\gamma} v^{[a} F^{b]c} \Bigg\} ~. \label{dotS1}
\end{eqnarray}
If the momenta had been defined in the usual way, the
$(\varphi-q/4\pi |r|^{3})$ terms would be absent from these
expressions. The extra complication in the present case derives from
the electromagnetic couplings in (\ref{pdefine}) and
(\ref{Sdefine}).

Unsurprisingly, (\ref{dotP1}) and (\ref{dotS1}) can be directly
related to the monopole and dipole moments of the force density
$F^{ab} J_{b}$. Denoting such moments by $\Psi^{a}(s)$ and
$\Psi^{ab}(s)$ respectively, it is possible to prove (\ref{fdefine})
and (\ref{tdefine}). It is then natural to refer to $-\Psi^{a}$ as
the net force, and $-2 \Psi^{[ab]}$ as the net torque acting on the
body. Viewing these quantities as force moments allows one to derive
(\ref{forcedefine}) and (\ref{torquedefine}). But these results are
no different than (\ref{dotP1}) and (\ref{dotS1}). If desired, the
reader may therefore view $\Psi^{a}$ and $\Psi^{[ab]}$ to be
\textit{defined} by (\ref{fdefine}) and (\ref{tdefine}).

In cases where $F^{ab}$ varies slowly in the center-of-mass frame
(both spatially and temporally), the expressions for the force and
torque can be expanded in Taylor series involving successively
higher multipole moments of the current density (denoted by
$Q^{\cdots}$). Deriving such equations would be awkward using the
methods introduced in this section, so we simply refer to
(\ref{forcemult}) and (\ref{torquemult}). Despite the peculiar
definitions of the current moments being used, these expansions are
exactly what one would expect out of a multipole formalism, and can
therefore be considered a check that Dixon's definitions are
reasonable.

In most cases where (\ref{forcemult}) and (\ref{torquemult}) are any
simpler than the exact expressions for the force and torque, only
the monopole and dipole moments will be significant. These moments
can be computed from (\ref{chargemono}) and (\ref{dipole})
respectively. In the limit that the particle is vastly smaller than
any of the field's length scales, only the monopole moment will
enter the equations of motion. In this case, the force reduces to
the standard Lorentz expression and the torque vanishes, as
expected.

The final ingredients required to complete this formalism are unique
prescriptions for $Z$ and $n^{a}$. Simply knowing the linear and
angular momenta at any point in time does not necessarily determine
the body's location in any useful way. The problem is compounded by
the fact that these quantities are strongly dependent on the choice
of reference frame itself. Indeed, without knowing where $W$ is, it
is essentially impossible to specify $J^{a}$ in any meaningful way.

These problems can be removed by choosing $Z$ and $n^{a}$
appropriately, and then assuming that the resulting worldline
provides a reasonable representation of the body's `average'
position. Following \cite{Dix70a,Dix74,Dix79,Ehl77}, we first assume
that for any point $z \in W$, there exists a unique future-directed
timelike unit vector $n^{a}(z)$ such that
\begin{equation}
p^{a}(z;n)=M(z;n) n^{a}(z) ~, \label{massdefine}
\end{equation}
for some positive scalar $M$. Here, we have temporarily changed the
dependencies of $p^{a}$ and $n^{a}$ for clarity. It is seen from
(\ref{pdefine}) that $p^{a}$ depends nontrivially on both the base
point $z$, and on $n^{a}$, which defines the surface of integration.
(\ref{massdefine}) is therefore a highly implicit definition of
$n^{a}$.

In any case, another condition must be also be given to fix $z$. We
want this to lie on a `center-of-mass line' in some sense, so it
would be reasonable to expect the `mass dipole moment' defined with
respect to it to vanish:
\begin{equation}
n_{b}n_{c} t^{abc} = 0 ~, \label{tempCM}
\end{equation}
where $t^{abc}$ is the full dipole moment of the stress-energy
tensor. This is defined by (\ref{stressdipole}), so (\ref{tempCM})
is equivalent to
\begin{equation}
n_{a}\left( z \right) S^{ab}\left( z ; n \right)=0 ~.
\label{CMdefine}
\end{equation}
The integral form of (\ref{CMdefine}) reduces to the standard
center-of-mass condition when $F^{ab}=0$ and $n^{a}=v^{a}$. It also
allows us to write the angular momentum in terms of a single
3-vector $S^{a}$
\begin{equation}
S^{ab} = \epsilon^{abcd} n_{c} S_{d} . \label{spinang}
\end{equation}

Of course, both of these properties would also have been satisfied
by instead requiring $v_{a} S^{ab} = 0$. We reject this choice due
to the fact that it leads to nonzero accelerations even when
$F^{ab}=0$ \cite{Mat}. Replacing $v^{a}$ by $n^{a}$ avoids this
peculiar behavior, which we consider to be an important requirement
for anything deserving to be called a center-of-mass line. It also
seems more natural to define the mass dipole moment in terms of
$n^{a}$ rather than $v^{a}$.

For the remainder of this paper, we assume that (\ref{massdefine})
and (\ref{CMdefine}) are always satisfied, and call the resulting
$Z$ and $n^{a}(s)$ the center of mass frame
\cite{Dix70a,Ehl77,Dix79}. It is not obvious that solutions to these
highly implicit equations exist, although existence and uniqueness
have been proven in the gravitational case \cite{Schatt1}. There, it
is also true that (given certain reasonable conditions) $Z$ is
necessarily a timelike worldline inside the convex hull of $W$. We
assume that the same results extend to electrodynamics.

The uniqueness of these definitions is actually not very critical
for our purposes. The important point is that \textit{a} solution
with the given properties can presumably be chosen for a
sufficiently large class of systems. While it is clearly very
difficult to find the center-of-mass directly from $T^{ab}$,
$J^{a}$, and $F^{ab}$, it is relatively straightforward to simply
construct sets of moments which automatically incorporate
(\ref{massdefine}) and (\ref{CMdefine}). (\ref{chargedensity}) and
(\ref{threecurrent}) show, for example, that these definitions do
not have any effect on our ability to construct arbitrary current
vectors adapted to them. There is nothing preventing the moment
potentials $\varphi$ and $H^{\alpha}$ from being appropriately
centered around $r^{\alpha}=0$. Although it is not obvious that this
can also be done for the stress-energy tensor, we conjecture that it
can.

Now that (\ref{massdefine}) and (\ref{CMdefine}) have been assumed
to hold, we need evolution equations for $M$, $n^{a}$, and $z^{a}$.
These are easily found from (\ref{fdefine}) and (\ref{tdefine}):
\begin{eqnarray}
\dot{M}&=&-n^{a}\Psi_{a} ~, \label{massevolve}
\\
M \dot{n}^{a}&=& -h^{a}_{b} \Psi^{b} ~, \label{nevolve}
\\
M(v^{a}-n^{a}) &=& S^{ab} \dot{n}_{b} - 2 \Psi^{[ab]} n_{b} ~,
\label{CMevolve}
\end{eqnarray}
where $h^{a}_{b}$ is the projection operator $\delta^{a}_{b}- n^{a}
n_{b}$. Note that the last of these equations shows that
$n^{a}=v^{a}$ if the spin and torque both vanish.

In general, (\ref{tdefine}) and (\ref{massevolve})-(\ref{CMevolve})
may be used together to find the motion of the body's center-of-mass
in terms of $\Psi^{a}$ and $\Psi^{[ab]}$. Once the field is known,
these quantities follow from (\ref{forcedefine}) and
(\ref{torquedefine}). Some recipe for evolving the dipole and higher
current moments in time -- most conveniently expressed in terms of
$\dot{\varphi}$ and $\dot{H}^{\alpha}$ -- is also required.
Combining all of these elements together leaves us with a
well-defined initial value problem that will determine $z^{a}$,
$n^{a}$, $M$, $S^{ab}$, and $J^{a}$. If the stress-energy tensor is
also desired, possible forms of it could in principle be constructed
from (\ref{stress1}) and (\ref{stress2}) in the same way that
$J^{a}$ was derived from $\varphi$ and $H^{\alpha}$. Combining all
of these steps would completely characterize the system, although we
shall omit the last one. The result is still sufficient to answer
most questions that one would be interested in asking of a nearly
isolated particle.

\section{The Field}
\label{FieldSect}

It is clear from the previous section that the equations of motion
will easily follow once $\Psi^{a}$ and $\Psi^{[ab]}$ are known.
These depend on the field, which we now calculate. For a reasonably
isolated body, it is first convenient to split $F^{ab}$ into two
parts
\begin{equation}
F^{ab} = F^{ab}_{\mathrm{ext}} + F^{ab}_{\mathrm{self}} ~.
\end{equation}
The external field is assumed to be generated by outside sources,
while the self-field is entirely due to the charge itself. Since
(\ref{forcedefine}) and (\ref{torquedefine}) are linear in $F^{ab}$,
the force and torque can also be split up into `self' and `external'
portions.

In Lorenz gauge, the vector potential sourced by the particle is
given by \cite{Jackson}
\begin{equation}
A^{a}_{\mathrm{self}}(x)=- \int \! \mathrm{d}^{4} x' \, J^{a}(x')
\delta \big( \sigma(x,x') \big) ~, \label{selfdefine}
\end{equation}
where
\begin{equation}
\sigma(x,x')=\frac{1}{2} (x-x')^{a}(x-x')_{a} ~, \label{sigdefine}
\end{equation}
is Synge's world function \cite{Synge,SFReview1}. In realistic
systems, this potential will couple to the external one via the
outside matter fields. These may act as reflectors or dielectrics,
or there may be an $n$-body interaction where the self-fields
influence the motion of some external charged particles (obviously
affecting the fields in $W$). Although it would be reasonable to
group together all portions of $F^{ab}$ causally related to our
particle in some way as the `self-field,' this would be impossible
to do with any generality. Instead, we simply define the
$F^{ab}_{\mathrm{self}}$ to be the field derived from
(\ref{selfdefine}) in the usual way (this differs from the point of
view taken in e.g. \cite{NewtMult}). The interactions of the
self-field with the outside world will all be categorized as
`external' effects that we presume can be accounted for by separate
methods.

If all of the external matter is sufficiently far away from $W$ (and
slowly varying), $F^{ab}_{\mathrm{ext}}$ will be approximately
constant within each $\Sigma(s) \cap W$ slice (and from one slice to
the `next'). $\Psi^{a}_{\mathrm{ext}}$ and
$\Psi^{[ab]}_{\mathrm{ext}}$ can therefore be approximated by
(\ref{forcemult}) and (\ref{torquemult}) in these cases. Finding the
self-force and self-torque is more complicated. For this,
$F^{ab}_{\mathrm{self}}$ has to be combined with the exact integral
expressions for the force and torque -- (\ref{forcedefine}) and
(\ref{torquedefine}). The detailed structure of the self-field must
therefore be known throughout $W$.

This easily follows from $A_{\mathrm{self}}^{a}$:
\begin{eqnarray}
F^{ab}_{\mathrm{self}}(x) &=& 2 \partial^{[a} A^{b]}_{\mathrm{self}}
~, \nonumber
\\
&=& - 2 \int \! \mathrm{d}^{4} x' \, \delta'(\sigma) (x-x')^{[a}
J^{b]} ~.
\end{eqnarray}
Writing $x$ in terms of $( r^{\alpha}, s )$, and $x'$ in terms of $(
r'^{\alpha}, \tau )$, and defining $\dot{\sigma}(x,x') :=
\partial \sigma/\partial \tau$, $F^{ab}_{\mathrm{self}}$ becomes
\begin{equation}
F^{ab}_{\mathrm{self} (\pm)}(x) = 2 \int \! \mathrm{d}^{3} r' \,
\Bigg\{ \frac{1}{|\dot{\sigma}|}
\frac{\mathrm{d}}{\mathrm{d}\tau}\left[ \frac{N}{\dot{\sigma}}
(x-x')^{[a} J^{b]} \right] \Bigg\}_{\tau= \tau_{\pm}} ~.
\label{field1}
\end{equation}
$\tau_{+}$ ($>s$) represents the advanced time, and $\tau_{-}$ the
retarded one. These are found by solving $\sigma(x,x')=0$ with $x$
and $r'$ held fixed.

Although we only consider the retarded field to be real, the
advanced solution is also retained for now. This allows us to later
construct the radiative self-field, which is considerably simpler
than the full retarded field. It would be quite convenient if the
self-forces generated by the two fields were identical (as Dirac
assumed for point particles \cite{Dirac}), although we will show
that this is not true in general.

Returning to the explicit form for $F^{ab}_{\mathrm{self}}$,
splitting up $J^{a}$ according to (\ref{Jtetrad}) shows that
\begin{eqnarray}
F^{ab}_{\mathrm{self}}(x) &=& 2 \int \! \mathrm{d}^{3} r'
\frac{N}{\dot{\sigma}|\dot{\sigma}|} \, \Bigg\{ \rho \left[
(x-x')^{[a} \dot{n}^{b]} - \left( \frac{\ddot{\sigma}}{ \dot{\sigma}
} + N^{-1} \ddot{n}^{\alpha} r'_{\alpha} \right) (x-x')^{[a} n^{b]}
- v^{[a} n^{b]} \right] + \dot{\rho} (x-x')^{[a} n^{b]} \nonumber
\\
&& {} - j^{\alpha} \left[ (x-x')^{[a} e^{b]}_{\alpha} \left(
\frac{\ddot{\sigma}}{ \dot{\sigma} } + N^{-1} \ddot{n}^{\beta}
r'_{\beta} \right) + \dot{n}_{\alpha} (x-x')^{[a} n^{b]} + \left( N
n^{[a} + (h \cdot v)^{[a} \right) e^{b]}_{\alpha} \right] \nonumber
\\
&& {} + \dot{j}^{\alpha} (x-x')^{[a} e^{b]}_{\alpha} \Bigg\}~.
\label{field2}
\end{eqnarray}
Here, $\dot{j}^{\alpha}(r',\tau) :=
\partial j^{\alpha}(r',\tau)/\partial \tau$, which differs from our
usual convention (e.g. $\dot{n}^{\alpha}(\tau) :=
e^{\alpha}_{a}(\tau) \mathrm{d}n^{a}(\tau)/\mathrm{d}\tau$).

Moving on, (\ref{sigdefine}) implies that
\begin{equation}
\dot{\sigma}(x,x') = - \bigg( N(r',\tau) n^{a}(\tau) + (h \cdot
v)^{a}(\tau) \bigg) (x-x')_{a} ~, \label{dotsig}
\end{equation}
and
\begin{equation}
\ddot{\sigma}(x,x') = N^{2} + v^{\alpha} v_{\alpha} - \left( N
\dot{n}^{a} - n^{a} \ddot{n}^{\beta} r'_{\beta} + \frac{\mathrm{d}(h
\cdot v)^{a}}{\mathrm{d}\tau} \right) (x-x')_{a}  ~.
\label{dotdotsig}
\end{equation}

If we now specify how $\varphi$ and $H^{\alpha}$ vary in time (which
determines $\dot{\rho}$ and $\dot{j}^{\alpha}$), we would have all
of the ingredients necessary to find the body's motion without any
approximation. Unfortunately, inserting (\ref{field2}) into
(\ref{tdefine}), (\ref{forcedefine}), (\ref{torquedefine}), and
(\ref{massevolve})-(\ref{CMevolve}) leads to a set of delay
integro-differential equations for the object's motion. Such a
system would be extremely difficult to solve (or even interpret) in
general, although it could be a useful starting point for numerical
simulations. It might also be interesting analytically if one were
looking for the forces required to make a body move in some
pre-determined way (e.g. a circular orbit).

Such questions will not be discussed here. We instead consider the
simplest possible approximations that allow us to gain insight into
a generic class of systems. In particular, it is assumed that all of
the quantities in (\ref{field2}) which depend on $\tau_{\pm}$ may be
written in terms of quantities at $s$ (via Taylor expansion). This
requires that nothing very drastic happen on timescales less than
the body's light-crossing time. This is not as trivial a condition
as it might seem to be (see e.g. \cite{Selfosc1}), although we will
not attempt to justify it.

Expressing everything in terms of $s$ rather than $\tau_{\pm}$ first
requires calculating $\Delta_{\pm} := s - \tau_{\pm}$. $\tau_{\pm}$
was defined by $\sigma =0$, so $\Delta_{\pm}$ can be found by Taylor
expanding this equation in $\Delta_{\pm}$. Assuming that both
$n^{a}$ and $v^{a}$ are at least $C^{3}$ in time for $s \in
[\tau_{\pm}, s]$,
\begin{eqnarray}
e^{a}_{\alpha}(\tau_{\pm}) &=& e^{a}_{\alpha}(s) + \Delta_{\pm}
n^{a}(s) \dot{n}_{\alpha}(s) - \frac{1}{2} \Delta_{\pm}^{2} \Big(
\dot{n}^{a}(s) \dot{n}_{\alpha}(s) + n^{a}(s) \ddot{n}_{\alpha}(s)
\Big) - \frac{1}{6} \Delta_{\pm}^{3}
\dddot{e}^{a}_{\alpha}\left(\xi_{\pm}^{(1)}\right) ~,
\label{expand1}
\\
z^{a}(\tau_{\pm}) &=& z^{a}(s) - \Delta_{\pm} v^{a}(s) + \frac{1}{2}
\Delta_{\pm}^{2} \dot{v}^{a}(s)  - \frac{1}{6} \Delta_{\pm}^{3}
\ddot{v}^{a}(s) + \frac{1}{24} \Delta_{\pm}^{4}
\dddot{v}^{a}\left(\xi_{\pm}^{(2)}\right) ~, \label{expand2}
\end{eqnarray}
where $\xi^{(1)}_{\pm}$ and $\xi^{(2)}_{\pm}$ are some numbers
between $\tau_{\pm}$ and $s$.

Then (\ref{XtoRS}) shows that
\begin{eqnarray}
(x-x')^{a} &\simeq& e^{a}_{\alpha} (r-r')^{\alpha} + \Delta_{\pm}
\Big( N n^{a} + (h \cdot v)^{a} \Big) + \frac{1}{2} \Delta_{\pm}^{2}
\Big( n^{a} \ddot{n}^{\alpha} r'_{\alpha} - N \dot{n}^{a} \Big) +
\frac{1}{6} \Delta_{\pm}^{3} \ddot{n}^{a} . \label{dX1}
\end{eqnarray}
Everything here has been written in terms of $s$, and all terms
involving quantities such as $|\dot{n}|^{3}$, $|\ddot{n}|^{2}$,
$|\dddot{n}|$, $|h \cdot v|^{2}$, and $|\dot{v}-\dot{n}|$ have been
removed. Since $|\Delta_{\pm}| \sim |r-r'|$, these terms can be
reasonably neglected if $|\dddot{n}| \mathcal{R}^{3} \ll |\ddot{n}|
\mathcal{R}^{2}  \ll |\dot{n}| \mathcal{R} \ll 1$, and
$|\dot{v}-\dot{n}| \mathcal{R} \ll |h \cdot v| \ll |\dot{n}|
\mathcal{R}$, where $\mathcal{R} := \max \big( |r|, |r'| \big)$. In
a sense, we are expanding up to second order in powers of
$|\dot{n}|\mathcal{R}$, and up to first order in $|v|$.

The requirement $|\dot{n}|\mathcal{R} \ll 1$ must hold for all
$(r,r')$ pairs, so it is useful to define the largest possible value
of $\mathcal{R}$. We call this the body's `radius'
\begin{equation}
D(s) := \max_{\Sigma(s) \cap W} |r| ~. \label{Ddefine}
\end{equation}
Using it, $|\dot{n}|\mathcal{R} \ll 1$ implies $|\dot{n}| D \ll 1$.
This technically restricts the allowable size of the charge,
although very few reasonable systems would actually be excluded.

Interpreting the relation satisfied by $(h \cdot v)^{a}$
($=v^{a}-n^{a}$) isn't quite as simple. Given (\ref{CMevolve}), it
is roughly equivalent to assuming that spin effects are present only
to lowest nontrivial order. This is not completely accurate, though,
and more precise statements will be given in the following section.

In any case, the inequalities following (\ref{dX1}) will be assumed
to hold from now on. Using them to expand $\sigma=0$, we find that
\begin{eqnarray}
R^{2} := |r-r'|^{2} &\simeq& \Delta_{\pm} \left[ 2 v^{\alpha}
(r-r')_{\alpha} + \Delta_{\pm} N(r,s) N(r',s) + \frac{1}{3}
\Delta^{2}_{\pm}  \ddot{n}^{\alpha} (r+2r')_{\alpha} + \frac{1}{12}
\Delta_{\pm}^{3} |\dot{n}|^{2} \right] ~.
\end{eqnarray}
All but the second term here is already `small,' but not quite
negligible in our approximation. The zeroth order expression for
$\Delta_{\pm}$ ($=\mp R$) may therefore be substituted into each of
these terms without any overall loss of accuracy. The resulting
equation is easily solved:
\begin{eqnarray}
\Delta_{\pm} &\simeq& \mp R \bigg[ 1 + \frac{1}{2} \dot{n}_{\alpha}
(r+r')^{\alpha} - \frac{1}{2} \dot{n}^{\alpha} \dot{n}^{\beta}
r_{\alpha} r'_{\beta} \pm \frac{1}{6} R \ddot{n}^{\alpha}
(r+2r')_{\alpha} + \frac{1}{24} R^{2} \dot{n}^{\alpha}
\dot{n}_{\alpha} + \frac{3}{8} \Big( \dot{n}^{\alpha}
(r+r')_{\alpha} \Big)^{2} \nonumber
\\
&& {} \pm R^{-1} v^{\alpha} (r-r')_{\alpha} \bigg] ~,
\label{Deltasolve1}
\end{eqnarray}
where everything is evaluated at $s$.

Although it would still be straightforward at this stage to compute
the exact error in (\ref{Deltasolve1}), it is not necessary.
Dimensional analysis shows that the neglected terms have magnitudes
$|\dddot{n}| \mathcal{R}^{4}$, $|\ddot{n}|^{2} \mathcal{R}^{5}$, $|
d(h \cdot v )/ds | \mathcal{R}^{2}$, $|d^{2}(h \cdot v)/ds^{2}|
\mathcal{R}^{3}$, $|h \cdot v|^{2} \mathcal{R}$, and so on (where
each of these can be evaluated anywhere in the interval
$[\tau_{\pm},s]$).

Continuing to expand quantities appearing in (\ref{field2}), a long
but straightforward calculation shows that (\ref{dotsig}) and
(\ref{dotdotsig}) can be approximated by
\begin{eqnarray}
\dot{\sigma} \simeq \pm R \left[ 1- \frac{1}{2} \dot{n}^{\alpha}
(r+r')_{\alpha} - \frac{1}{8} \Big( \dot{n}^{\alpha} (r-r')_{\alpha}
\Big)^{2} \mp \frac{1}{3} R \ddot{n}^{\alpha} (r+2r')_{\alpha} +
\frac{1}{8} R^{2} |\dot{n}|^{2} \right] ~,
\end{eqnarray}
and
\begin{eqnarray}
\ddot{\sigma} \simeq 1 - \dot{n}^{\alpha} (r+r')_{\alpha} \mp R
\ddot{n}^{\alpha} (r+2r')_{\alpha} + \dot{n}^{\alpha}
\dot{n}^{\beta} r_{\alpha} r'_{\beta} + \frac{1}{2} R^{2}
|\dot{n}|^{2} ~.
\end{eqnarray}

Looking at (\ref{field2}), the charge and current densities are the
only quantities that have not yet been expanded away from
$\tau_{\pm}$. It is instructive to temporarily leave them like this,
but write out all other terms in field at $s$. The resulting
expression is quite long, so we break it up into several smaller
pieces by writing $F^{ab}_{\mathrm{self}(\pm)}$ in the form
\begin{equation}
F^{ab}_{\mathrm{self}(\pm)} = 2 \int \! \mathrm{d}^{3} r'
\frac{1}{R^{3}} \, \bigg( \rho(r',\tau_{\pm}) f^{ab}_{(1)} +
\dot{\rho}(r',\tau_{\pm}) f^{ab}_{(2)} - j^{\beta} (r',\tau_{\pm})
f^{ab}_{\beta(3)} + \dot{j}^{\beta}(r',\tau_{\pm}) f^{ab}_{\beta(4)}
\bigg) ~. \label{field3}
\end{equation}
The definitions of each of these coefficients is obvious from
comparison with (\ref{field2}).

Before computing them, we first simplify the notation by defining a
quantity $T(s)>0$ such that $T^{-1}$ remains (marginally) less than
$|\dot{n}|$, $|\ddot{n}|^{1/2}$, $|h \cdot v|^{1/2}/D$, etc. This is
useful because a number of different objects were assumed to be
negligible in this section. Some of these may be much larger than
the others, so using them to writing down error estimates would
become rather awkward. Introducing $T$ removes this difficulty.

Along the same lines, we also define $\epsilon := D/T \ll 1$. A
simple but exceedingly tedious calculation then shows that the
coefficients in (\ref{field3}) are equal to
\begin{eqnarray}
f^{ab}_{(1)} &\simeq& - (r-r')^{\alpha} e_{\alpha}^{[a} n^{b]}
\left[ 1+ \frac{1}{2} \dot{n}^{\beta}(r-r')_{\beta} - \frac{1}{8}
\Big( \dot{n}^{\beta} (r-r')_{\beta} \Big)^{2} + \frac{1}{2}
\dot{n}^{\beta} \dot{n}^{\gamma} r_{\beta} (r-r')_{\gamma}  +
\frac{1}{8} R^{2} | \dot{n} |^{2} \right] \nonumber
\\
&& {} + \frac{1}{2} R^{2} (r-r')^{\alpha} e_{\alpha}^{[a}
\ddot{n}^{b]} - \frac{1}{2} R^{2} n^{[a} \dot{n}^{b]} \left( 1 +
\frac{1}{2} \dot{n}^{\beta} (3r-r')_{\beta} \right) \pm \frac{2}{3}
R^{3} \ddot{n}^{[a} n^{b]} + o \left(  \Big[\epsilon (\mathcal{R}/
D) \Big]^{3} \mathcal{R} \right) ~, \label{f1}
\\
f^{ab}_{(2)} &\simeq& \pm R \, \Bigg\{ (r-r')^{\alpha}
e_{\alpha}^{[a} n^{b]} \left[ 1 + \dot{n}^{\beta} r_{\beta} + \Big(
\dot{n}^{\beta} r_{\beta} \Big)^{2} \pm \frac{1}{3} R
\ddot{n}^{\beta} (2r+r')_{\beta} - \frac{1}{4} R^{2} |\dot{n}|^{2}
\right] \pm \frac{1}{2} R^{2} (r-r')^{\alpha} e_{\alpha}^{[a}
\ddot{n}^{b]} \nonumber
\\
&&  {} + (r-r')^{\alpha} e_{\alpha}^{[a} \dot{n}^{b]} R \left( 1 +
\frac{1}{2} \dot{n}^{\beta} (3r+r')_{\beta} \right) \mp \frac{1}{2}
R^{2} n^{[a} \dot{n}^{b]} \left( 1 + 2 \dot{n}^{\beta} r_{\beta}
\right) - \frac{1}{3} n^{[a} \ddot{n}^{b]} R^{3} - (h \cdot v)^{[a}
n^{b]} R \Bigg\} \nonumber
\\
&& {} + o \left(  \Big[\epsilon (\mathcal{R}/ D) \Big]^{3}
\mathcal{R}^{2} \right) ~, \label{f2}
\\
f^{ab}_{\beta(3)} &\simeq& (r-r')^{\alpha} e_{\alpha}^{[a}
e_{\beta}^{b]} \left[ 1 + \frac{1}{2} \dot{n}^{\gamma}
(r-r')_{\gamma} - \frac{1}{8} \Big( \dot{n}^{\gamma} (r-r')_{\gamma}
\Big)^{2} + \frac{1}{2} \dot{n}^{\gamma} \dot{n}^{\lambda}
r_{\gamma} (r-r')_{\lambda} + \frac{1}{8} R^{2} |\dot{n}|^{2}
\right] \nonumber
\\
&& {} + \frac{1}{2} R^{2} \ddot{n}_{\beta} (r-r')^{\alpha}
e_{\alpha}^{[a} n^{b]} + \frac{1}{2} R^{2} \dot{n}_{\beta}
(r-r')^{\alpha} e_{\alpha}^{[a} \dot{n}^{b]} + n^{[a} e^{b]}_{\beta}
R^{2} \bigg( \frac{1}{2} \ddot{n}^{\gamma} (r-r')_{\gamma} \mp
\frac{1}{3} R |\dot{n}|^{2} \nonumber
\\
&& {} - R^{-2} v^{\gamma} (r-r')_{\gamma} \bigg) + \frac{1}{2} R^{2}
\dot{n}^{[a} e^{b]}_{\beta} \left( 1 + \frac{1}{2} \dot{n}^{\gamma}
(3r-r')_{\gamma} \right) \pm \frac{1}{3} R^{3} \ddot{n}^{[a}
e^{b]}_{\beta} + o \left(  \Big[\epsilon (\mathcal{R}/ D) \Big]^{3}
\mathcal{R} \right) ~, \label{f3}
\\
f^{ab}_{\beta(4)} &\simeq& \pm R \, \Bigg\{ (r-r')^{\alpha}
e_{\alpha}^{[a} e_{\beta}^{b]} \left[ 1 + \dot{n}^{\beta} r_{\beta}
+ \Big( \dot{n}^{\beta} r_{\beta} \Big)^{2} \pm \frac{1}{3} R
\ddot{n}^{\beta} (2r+r')_{\beta} - \frac{1}{4} R^{2} |\dot{n}|^{2}
\right] \mp \frac{1}{2} R^{2} (r-r')^{\alpha} e_{\alpha}^{[a}
\nonumber
\\
&& {} \times \big( \dot{n}^{b]} \dot{n}_{\beta} -
n^{b]}\ddot{n}_{\beta} \big) - (r-r')^{\alpha} e_{\alpha}^{[a}
n^{b]} \dot{n}_{\beta} R \left( 1 + \frac{1}{2} \dot{n}^{\gamma}
(3r+r')_{\gamma} \right)  - n^{[a} e^{b]}_{\beta} R \left[ 1+
\frac{1}{2} \dot{n}^{\gamma} (3r-r')_{\gamma} \right. \nonumber
\\
&& \left. {} - \frac{1}{8} \Big( \dot{n}^{\gamma} (r+r')^{\gamma}
\Big)^{2} + \dot{n}^{\gamma} \dot{n}^{\lambda} r_{\gamma} (2
r_{\lambda} + r'_{\lambda} ) \pm \frac{1}{6} R \ddot{n}^{\gamma} (5
r + r')_{\gamma} - \frac{7}{24} R^{2} |\dot{n}|^{2} \pm R^{-1}
v^{\gamma} (r-r')_{\gamma} \right] \nonumber
\\
&& {} - (h \cdot v)^{[a} e^{b]}_{\beta} R \mp \frac{1}{2}
\dot{n}^{[a} e^{b]}_{\beta} R^{2} \left( 1+2 \dot{n}^{\gamma}
r_{\gamma} \right) - \frac{1}{6} R^{3} \ddot{n}^{[a} e^{b]}_{\beta}
\Bigg\} + o\left(  \Big[\epsilon (\mathcal{R}/ D) \Big]^{3}
\mathcal{R}^{2} \right) ~. \label{f4}
\end{eqnarray}

\subsection{Approximations}
\label{Approximations}

At this point, it is useful to consider the physical meaning behind
our approximations more carefully. Although error estimates have
been given for for $f^{ab}_{(1)}$-$f^{ab}_{\beta (4)}$, the
important errors are those in $F^{ab}_{\mathrm{self}}$.
Unfortunately, the unbounded integrand in (\ref{field3}) makes it
difficult to compute these rigorously. We shall simply assume that
$J^{a}$ is sufficiently homogeneous that dimensional analysis can be
used to say, for example, that the absolute value of the error in
$\int \! \mathrm{d}^{3} r' \, \rho(r',\tau_{\pm})
f_{(1)}^{ab}/R^{3}$ is less than about
\begin{equation}
\epsilon^{3} \max |\rho| D ~.
\end{equation}

With this established, we can also use the assumptions outlined in
the previous section to provide bounds on the force and torque.
Combining  $|\dot{n}| < T^{-1}$ with (\ref{nevolve}),
\begin{eqnarray}
\left| \Psi^{\alpha} \right| < M T^{-1} ~. \label{eforcerestrict}
\end{eqnarray}
Since $T \gg D$, this means that the particle cannot be accelerated
up to an appreciable fraction of the speed of light within a
light-crossing time -- a very reasonable restriction.

Differentiating (\ref{nevolve}),
\begin{equation}
M \ddot{n}^{\alpha} = - \dot{\Psi}^{\alpha} - 2 \dot{M}
\dot{n}^{\alpha} ~.
\end{equation}
Generically, this implies that $|\ddot{n}| < T^{-2}$ can be ensured
by assuming
\begin{eqnarray}
\left|\dot{M}\right| &<& M T^{-1} ~, \label{nforcerestrict}
\\
\left| \dot{\Psi}^{\alpha} \right| &<& M T^{-2} ~.
\end{eqnarray}

Differentiating (\ref{nevolve}) a second time and using
(\ref{massevolve}), we also have that
\begin{eqnarray}
\left| \ddot{M} \right| &<& M T^{-2} ~, \label{massrestrict1}
\\
\left| \ddot{\Psi}^{\alpha} \right| &<& M T^{-3} ~,
\\
\left| n_{a} \dot{\Psi}^{a} \right| &<& M T^{-2} ~.
\end{eqnarray}

Similar restrictions on the spin and torque can be found by using
(\ref{CMevolve}). In particular, $|v|$ will remain less than
$\epsilon^{2}$ if
\begin{eqnarray}
\left|S^{\alpha \beta}\right| := \left|S^{\alpha}\right| &<&
\epsilon MD ~, \label{spinrestrict}
\\
\left| \Psi^{[ab]} e_{a}^{\alpha} n_{b} \right| &<& \epsilon^{2} M
~. \label{torquerestrict2}
\end{eqnarray}

Differentiating (\ref{CMevolve}), $|\dot{n}-\dot{v}| D \ll
\epsilon^{2}$ is implied by
\begin{eqnarray}
\left| \Psi^{[\alpha \beta]} \right| &\ll& \epsilon M ~,
\label{torquerestrict}
\\
\left| \dot{\Psi}^{[ab]} e_{a}^{\alpha} n_{b} \right| &\ll& \epsilon
MD  ~.
\end{eqnarray}

Although weaker restrictions than these can be adopted in special
cases, we assume for simplicity that they always hold. More
concisely, we are restricting ourselves to systems where the
magnitude of each tetrad component of the (full self + external)
force is bounded by about $M T^{-1}$, and the magnitude of each
torque component is no larger than $\epsilon^{2} M$. Bounds on the
$s$-derivatives of these quantities are suppressed by appropriate
factors of $T$. These conditions basically mean that the body can
only change significantly over timescales larger than $T$.
(\ref{spinrestrict}), for example, implies that this timescale sets
a lower bound on the charge's rotational period.

These interpretations should not be surprising. Unfortunately,
though, they are quite not as simple as they appear. Combining them
with (\ref{forcedefine}) and (\ref{torquedefine}) can lead to less
obvious bounds on the structure of $J^{a}$ itself. One of these
affects the size of the particle's dipole moment. If the self-torque
is small compared to the external one, then the magnitudes of
\begin{equation}
Q^{[\alpha}{}_{\gamma} F^{\beta] \gamma}_{\mathrm{ext}}, \, \, n_{a}
e^{c}_{\gamma} Q^a{}_{c} F^{\beta \gamma}_{\mathrm{ext}} , \, \,
n_{a} e_{c}^{\gamma} Q^\beta{}_{\gamma} F^{ac}_{\mathrm{ext}}, \, \,
e_{a}^{[\alpha} e_{b}^{\beta]} n_{c} n_{d} Q^{ac}
F^{bd}_{\mathrm{ext}} \label{dipoleBound}
\end{equation}
should all be less than about $\epsilon^{2}M$. If $q$ is negligible,
these conditions ensure that both the force and torque are
sufficiently small.

In most cases of interest, though, $q \neq 0$. Bounding the
quantities in (\ref{dipoleBound}) then guarantees only that the
torque is acceptable. The (full) force will often be dominated by
the monopole component of $\Psi^{a}_{\mathrm{ext}}$, in which case
(applying (\ref{eforcerestrict}) and (\ref{nforcerestrict}))
\begin{eqnarray}
M T^{-1} &>& \left| q e_{a}^{\alpha} n_{b} F^{ab}_{\mathrm{ext}}
\right| ~,
\\
&>& \left| q v_{\beta} F^{\alpha \beta}_{\mathrm{ext}} \right| ~.
\end{eqnarray}
Since $v_{\beta}$ is small, this second bound allows the magnetic
field to be quite large. Despite this, we choose to assume (purely
for simplicity) that both the electric and magnetic fields are
bounded by $M\big(|q|T\big)^{-1}$. If the fields are in fact as
large as these equations allow, then the previous bounds on the
dipole moment imply that
\begin{equation}
\left|Q^{\alpha \beta}\right| \lesssim \epsilon |q| D~,
\label{dipolerestrict}
\end{equation}
along with a similar restriction on $\left|Q^{ab} e_{a}^{\alpha}
n_{b}\right|$. This implies that the `center-of-charge' must be very
close to the center-of-mass, which considerably reduces the number
of current distributions that we can allow. Note, however, that
these assumptions could be relaxed somewhat if extra conditions are
placed on the sizes of the field components.

To find some other implications of our approximations, we now study
a particular example in detail before moving on to computing general
self-forces.

\section{Self-Forces}
\label{ForceSect}

\subsection{Stationary Case}

It was shown in the previous section that (\ref{field3})-(\ref{f4})
approximate the self-field due to a charge with an (almost)
arbitrarily accelerating center-of-mass line. These equations become
exact when $\dot{n}^{a} =0$ and $n^{a}=v^{a}$. Temporarily assuming
that these conditions are true,
\begin{eqnarray}
F^{ab}_{\mathrm{self}(\pm)}(x) &=& - 2 \int \! \mathrm{d}^{3} r'
\frac{1}{R^{3}} \, \bigg[ (r-r')^{\alpha} e_{\alpha}^{[a} n^{b]}
\Big(  \rho(r',\tau_{\pm}) \mp R \dot{\rho}(r', \tau_{\pm}) \Big) -
(r-r')^{\alpha} e^{[a}_{\alpha} e^{b]}_{\beta} \Big( j^{\beta}(r',
\tau_{\pm}) \nonumber
\\
&& {} \pm R \dot{j}^{\beta} (r',\tau_{\pm}) \Big) + n^{[a}
e^{b]}_{\beta} R^{2} \dot{j}^{\beta}(r',\tau_{\pm}) \bigg] ~,
\label{statfield}
\end{eqnarray}
where $\tau_{\pm} = s \pm R$.

One might at first think that the self-forces generated by this
field would be identically zero, but this is not generally correct.
Anything emitting a focused beam of radiation, for example, will
experience some recoil. It will also lose a bit of mass over time,
and can even start rotating if the the beam is offset from the
source's center-of-mass. These can all be thought of as self-force
and self-torque effects, and clearly do not require $\dot{v}^{a}
\neq 0$.

Since (\ref{statfield}) is exact, one could compute self-forces and
self-torques for a variety of systems without any approximation. If
these were nonzero, the assumptions leading to (\ref{statfield})
would not be maintained unless there were also external forces and
torques present that exactly balanced them. Such a situation would
be rather artificial, so we instead apply the approximations of Sec.
\ref{FieldSect}, and use the resulting self-force and self-torque to
gain some intuition into more general cases.

Assuming that $| \varphi | \gg |\dot{\varphi}| D \gg
|\ddot{\varphi}| D^{2}$, except perhaps at isolated points, one can
show that
\begin{eqnarray}
\Psi^{\alpha}_{\mathrm{self}(\pm)} &\simeq& \int \! \mathrm{d}^{3} r
\, \mathrm{d}^{3} r' \, \frac{1}{R^{3}} \, \Bigg\{ R^{2} \rho(r,s)
\dot{H}^{\alpha}(r',s)   - 2 (r-r)^{\alpha} H_{\beta}(r,s) \left(
\frac{1}{2} r'^{\beta} \dot{\varphi}(r',s) \mp R
\dot{H}^{\beta}(r',s) \right) \nonumber
\\
&&  {} + \left( \varphi(r,s) - \frac{q}{4\pi |r|^{3}} \right) \bigg[
(r-r')^{\alpha} r_{\beta} - \delta^{\alpha}_{\beta} (r-r')^{\gamma}
r_{\gamma} \bigg] \dot{H}^{\beta}(r',s) \Bigg\} ~, \label{ePsi2}
\end{eqnarray}
to first order in $\dot{\varphi}$ and $\dot{H}^{\alpha}$. This
illustrates that $\Psi^{\alpha}$ is bounded by $\dot{\varphi} H
D^{5}$ (meaning $\max|\dot{\varphi}| \max|H| D^{5}$), $\varphi
\dot{H} D^{5}$, or $H \dot{H} D^{5}$ (whichever is largest). If
$H^{\alpha}$ is very small compared to $\varphi$, then the lowest
order contributions to the self-force can be shown to be of order
$\varphi \ddot{\varphi} D^{6}$.

It is then clear that unless the system is particularly symmetric,
self-forces with these magnitudes are unavoidable. They would
considerably complicate the equations of motion when considering
nonzero center-of-mass accelerations, so it would be convenient to
ignore them. This is self-consistent with the approximations already
in place if the charge's `elasticity' is bounded by
\begin{eqnarray}
\dot{\varphi} D &<& \epsilon^{2} \varphi ~, \label{phisize}
\\
\ddot{\varphi} D^{2} &\ll& \epsilon^{2} \varphi ~, \label{phisize2}
\\
\dot{H}  D &\ll& \epsilon^{2} H ~. \label{Hsize}
\end{eqnarray}

If these relations hold, the self-force becomes negligible whenever
$\dot{n}^{a}=0$ and $n^{a}=v^{a}$; i.e. a charge at rest will remain
at rest unless acted on by an external field. For this reason, all
of the calculations that follow will assume (\ref{phisize}) and
(\ref{phisize2}) to be true. (\ref{Hsize}) is extremely simple to
relax, so it will be eventually be replaced with $\dot{H} D <
\epsilon^{2} H$. The interested reader could easily weaken these
conditions even further, although we regard the extra complication
to be unnecessary for the present purposes.

In writing down (\ref{statfield}), both $\dot{n}^{a}=0$ and
$n^{a}=v^{a}$ were required. This second condition was adopted here
for simplicity, although it is by no means `natural.' Given
(\ref{CMevolve}), it can only hold if $\Psi^{[ab]} n_{b}=0$. But
combining (\ref{phisize})-(\ref{Hsize}) with (\ref{torquedefine})
and (\ref{statfield}) shows that this portion of the torque is
usually nonzero even if $\dot{\varphi}=\dot{H}^{\alpha}=0$.
Generically, one finds that $\Psi^{[ab]} n_{b}$ will only be small
if either $\varphi \ll H$ or $H \ll \varphi$. $n^{a} \simeq v^{a}$
is not a particularly important requirement, so we choose not to
impose these restrictions on the current structure. In general,
then, $n^{a} \neq v^{a}$ even when $\dot{n}^{a}=0$ (and $\varphi$
can be of order $H$).

Of course, $|n-v|=|v|$ still can't exceed $\epsilon^{2}$. As was
already shown, this remains true if each tetrad component of
$\Psi^{[ab]}$ is bounded by $\epsilon^{2} M$ (and $|S| < \epsilon
MD$). We now strengthen this assumption slightly, and require the
self and external components of the torque to be
\textit{individually} less than $\epsilon^{2} M$. There are then
terms in the self-torque which violate this unless more restrictions
are placed on the particle's current structure.

Using (\ref{statfield}) and (\ref{torquedefine}) again, one can see
that the tetrad components of $\Psi^{[ab]}_{\mathrm{self}}$ will be
of order $\varphi H D^{5}$, $\varphi \dot{\varphi} D^{6}$, etc.
These types of quantities can be conveniently simplified by defining
an `electromagnetic radius' $D_{\mathrm{em}} \sim \left( \varphi^{2}
+ H^{2} \right) D^{6}/M \lesssim D$. $D_{\mathrm{em}}/D$ then
estimates the fraction of the particle's mass that is of
(macroscopic) electromagnetic origin. This ratio will come up often,
so let
\begin{equation}
\mathcal{E} := \left( \frac{D_{\mathrm{em}}}{D} \right) \lesssim 1
~.
\end{equation}

In this notation, $\varphi \dot{\varphi} D^{6} \rightarrow
\epsilon^{2} \mathcal{E} M$, and $\varphi H D^{5} \rightarrow
\mathcal{E} M$. If $\varphi$ and $H$ are comparable, the second of
these expressions is clearly the one that estimates the self-torque.
Enforcing (\ref{torquerestrict2}) in this case therefore requires
\begin{equation}
\mathcal{E} < \epsilon^{2} ~. \label{selfenergyrestrict}
\end{equation}
Conditions weaker than this can be adopted if the particle is
dominated by either $\varphi$ or $H^{\alpha}$, although it will
still be true that $\mathcal{E} \ll 1$ (except in special cases).
Surprisingly, our approximations place a severe restriction on the
charge's self-energy. This makes it impossible to ever take the
point particle limit in a strict sense (if such a thing is even
meaningful). Such a procedure is not, however, necessary to answer
whether a particle's size is important when it is much smaller than
the characteristic scales of the surrounding system (the `physical
point particle limit').

It is interesting to note that this condition is really independent
of any of Dixon's special constructions. In almost any slow motion
approximation, one would expect that $|\dot{S}| \lesssim |S| T^{-1}$
and $|S|\lesssim MD^{2} T^{-1}$. So $\left|\Psi^{[ab]}\right| \sim
|\dot{S}| \lesssim \epsilon^{2} M $ is generic. And this is exactly
the condition that led to (\ref{selfenergyrestrict}). The specific
form of the torque needed to show this did use Dixon's definition,
but the more common one wouldn't have changed anything. This can be
verified by substituting (\ref{statfield}) into the first term in
(\ref{torquedefine}).

Also note that when $\mathcal{E} \sim 1$, it is not clear that any
slow motion assumption is even physically reasonable. It seems
difficult, for example, to avoid the extremely high frequency
self-sustaining oscillations discussed by Bohm and Weinstein
\cite{Selfosc1}, among others. When $\mathcal{E}$ is small, it is
easy to imagine that oscillations like these would have extremely
small amplitudes, or at least be highly damped in causal systems.
But the degree of rigidity required to hold together a charge with
very large self-energy might prevent this.

\subsection{Arbitrary Motion}

We now turn back to analyzing the self-forces and self-torques when
$\dot{n}^{a} \neq 0$ and $n^{a} \neq v^{a}$. For simplicity, it is
assumed that the internal dynamics are always `slow' in the sense of
(\ref{phisize}) and (\ref{phisize2}). It is simple enough to relax
(\ref{Hsize}), so we also allow the first (and only the first)
derivative of $H^{\alpha}$ to appear in the self-force. Higher
derivatives of both $\varphi$ and $H^{\alpha}$ could be included
with relatively little extra effort, although there seems to be
little reason for doing so.

\subsubsection{Radiative Self-Forces}
\label{RadiativeSect}

At this point, it is useful to separately write out the retarded and
radiative fields. Using (\ref{radfielddefine}) and
(\ref{field3})-(\ref{f4}),
\begin{equation}
F^{ab}_{\mathrm{self}(R)}(x) \simeq - \frac{4}{3} q \ddot{n}^{[a}(s)
n^{b]}(s) ~, \label{radfield}
\end{equation}
If $n^{a}=v^{a}$, then this is the same expression found by Dirac
from the Li\'{e}nard-Wiechert potential \cite{Dirac} (we have used
the opposite sign convention for the field).

The forces and torques exerted by this field are now very simple to
calculate. Inserting (\ref{radfield}) in (\ref{forcedefine}),
\begin{equation}
\Psi^{\alpha}_{\mathrm{self}(R)}(s) \simeq - \frac{2}{3} q^{2}
\ddot{n}^{\alpha} + o\big( \epsilon^{2} \mathcal{E} M T^{-1} \big)
~.
\end{equation}
This is just the Lorentz-Dirac force when $n^{a}=v^{a}$ (see
(\ref{LD})).

Substituting (\ref{radfield}) into (\ref{torquedefine}), the
space-space components of the self-torque are
\begin{equation}
\Psi^{[\alpha \beta]}_{\mathrm{self}(R)}(s) \simeq  \frac{1}{q} \!
\int \! \mathrm{d}^{3} r \, \rho(r,s) r^{[\alpha}
\Psi^{\beta]}_{\mathrm{self}(R)} + o \big( \epsilon^{3} \mathcal{E}
M \big) ~.
\end{equation}
$\int \! \mathrm{d}^{3} r \, \rho r^{\alpha}/q$ can be thought of as
the separation vector between the `charge centroid' and the
center-of-mass ($r^{\alpha}=0$). Inverting (\ref{spinang}), we can
clearly convert this component of the self-torque into a 3-vector
describing the rate of change of $S^{a}$. In this form, the (vector)
self-torque is just the cross product of the self-force with this
separation vector. Writing it in this way suggests that it arises
due to the self-force not acting through the center-of-mass.

The time-space components of the self-torque are similar:
\begin{equation}
\Psi^{[ab]}_{\mathrm{self}(R)}(s) n_{a} e_{b}^{\beta} \simeq
\frac{1}{2q} \! \int \! \mathrm{d}^{3} r \, H_{\alpha}(r,s)
\Psi^{\alpha}_{(R)\mathrm{self}} r^{\beta} + o \big( \epsilon^{3}
\mathcal{E} M \big)~. \label{ntorquerad}
\end{equation}
This clearly vanishes if the `current centroid' coincides with the
center-of-mass.

Combining all of these results with (\ref{tdefine}),
(\ref{massevolve}), (\ref{nevolve}), and (\ref{CMevolve}), one can
find that the particle's motion by simultaneously solving
\begin{eqnarray}
\dot{M} &\simeq& -n_{a} \Psi^{a}_{\mathrm{ext}} + o\big(
\epsilon^{2} \mathcal{E} M T^{-1} \big) ~, \label{dotmrad1}
\\
\dot{S}^{\alpha \beta} &\simeq& - 2 \Psi^{[\alpha
\beta]}_{\mathrm{ext}} + \frac{4}{3} q \! \int \! \mathrm{d}^{3} r
\, \rho(r,s) r^{[\alpha} \ddot{n}^{\beta]} + o\big( \epsilon^{3}
\mathcal{E} M \big) ~, \label{torquerad1}
\\
M \dot{n}^{\alpha} &\simeq& -\Psi^{\alpha}_{\mathrm{ext}} +
\frac{2}{3} q^{2} \ddot{n}^{\alpha} + o\big( \epsilon^{2}
\mathcal{E} M T^{-1} \big) ~, \label{forcerad1}
\\
M \dot{v}^{\alpha} &\simeq& M \dot{n}^{\alpha} + S^{\alpha \beta}
\ddot{n}_{\beta} - \dot{M} v^{\alpha} -4 \Psi^{ [\alpha \beta]
}_{\mathrm{ext}} \dot{n}_{\beta} - 2
\dot{\Psi}^{[ab]}_{\mathrm{ext}} e_{a}^{\alpha} n_{b} + o\big(
\epsilon^{2} \mathcal{E} M T^{-1} \big) ~. \label{forcerad2}
\end{eqnarray}

If the external field varies slowly over $\Sigma(s) \cap W$,
(\ref{forcemult}) and (\ref{torquemult}) can be used to approximate
$\Psi^{a}_{\mathrm{ext}}$ and $\Psi^{[ab]}_{\mathrm{ext}}$. Let the
minimum characteristic length scale of the external field be denoted
by $\lambda \gg D$, so that $|F^{ab}_{\mathrm{ext}}| \gtrsim \lambda
|\partial F^{ab}_{\mathrm{ext}}|$ (where the absolute value signs
are meant to act on each tetrad component of the quantity inside
them). It is then convenient to assume that
\begin{equation}
\lambda \gtrsim T/\mathcal{E} ~. \label{lambdarestrict}
\end{equation}
This ensures such that the dipole and higher contributions to the
external force are negligible compared to the Lorentz-Dirac
self-force.

Without any loss of accuracy, (\ref{lambdarestrict}) allows
(\ref{ntorquerad})-(\ref{forcerad2}) to be written as (making some
weak assumptions on the magnitudes of the quadrupole and higher
moments)
\begin{eqnarray}
\dot{M} &\simeq& -q n_{a}  e_{b}^{\beta} v_{\beta}
F^{ab}_{\mathrm{ext}}  ~, \label{dotmrad2}
\\
\dot{S}^{\alpha \beta} &\simeq& -2 e_{a}^{\alpha} e_{b}^{\beta}
\left( Q^{[a}{}_{c} F^{b] c}_{\mathrm{ext}} + Q^{d[a}{}_{c}
\partial_{d} F^{b] c}_{\mathrm{ext}} \right) + \frac{4}{3} q \! \int
\! \mathrm{d}^{3} r \, \rho(r,s) r^{[\alpha} \ddot{n}^{\beta]} ~,
\label{torquerad2}
\\
M \dot{n}^{\alpha} &\simeq& - q e_{a}^{\alpha} v_{b}
F^{ab}_{\mathrm{ext}}  + \frac{2}{3} q^{2} \ddot{n}^{\alpha} ~,
\label{forcerad3}
\\
M \dot{v}^{\alpha} &\simeq& M \dot{n}^{\alpha} + S^{\alpha \beta}
\ddot{n}_{\beta} - \dot{M} v^{\alpha} - 4 e_{a}^{\alpha}
e_{b}^{\beta} \dot{n}_{\beta} Q^{[a}{}_{c} F^{b]c}_{\mathrm{ext}} -
2 e^{\alpha}_{a} n_{b} \dot{Q}^{[a}{}_{c} F^{b]c}_{\mathrm{ext}}  ~,
\label{forcerad4}
\end{eqnarray}
where the error terms haven't changed. Similarly, $v^{\alpha}$ can
be recovered from (\ref{CMevolve}):
\begin{equation}
M v^{\alpha} \simeq S^{\alpha \beta} \dot{n}_{\beta} - 2
e_{a}^{\alpha} n_{b} \left( Q^{[a}{}_{c} F^{b] c}_{\mathrm{ext}} +
Q^{d[a}{}_{c} \partial_{d} F^{b] c}_{\mathrm{ext}} \right)  -
\frac{2}{3} q \! \int \! \mathrm{d}^{3} r \, H_{\beta}(r,s)
\ddot{n}^{\beta} r^{\alpha} + o \big( \epsilon^{3} \mathcal{E} M
\big) ~. \label{vrad1}
\end{equation}

The assumptions we've made so far -- although fairly restrictive --
are clearly not sufficient to recover the standard point particle
result. Still, there must exist some class of charges which do
behave in this way, and it is interesting to characterize it. Before
doing so, it must first be mentioned that the $\bar{s}$ in
(\ref{LD}) is a proper time, while $s$ is not. It is close, though.
Letting $s= s(\bar{s})$ and $s'(\bar{s}) := \mathrm{d}s/\mathrm{d}
\bar{s}$,
\begin{eqnarray}
s' &=& \frac{1}{\sqrt{v^{a}v_{a}}} ~, \label{StoSbar}
\\
&\sim& 1 + o\left(\epsilon^{4}\right) ~.
\end{eqnarray}

The fractional difference between the center-of-mass 4-velocity and
$v^{a}$ is therefore of order $\epsilon^{4}$. In particular, the
triad components of this 4-velocity will differ from $v^{\alpha}$ by
terms no larger than $\epsilon^{6}$. Given the error estimate in
(\ref{vrad1}), these differences are negligible whenever
$\mathcal{E} \gtrsim \epsilon^{3}$.

The proper acceleration can now be written in terms of $v^{a}$ and
$\dot{v}^{a}$:
\begin{equation}
z''^{a} = s'' v^{a} + \left( s' \right)^{2} \dot{v}^{a} ~.
\end{equation}
It follows from (\ref{forcerad4}) and (\ref{StoSbar}) that $s'' \sim
\epsilon^{4}/T$, so $z''^{a}$ differs from $\dot{v}^{a}$ by terms of
this same order. (\ref{forcerad2}) then implies that these two
quantities are interchangeable whenever $\mathcal{E} \gtrsim
\epsilon^{2}$.

Applying similar arguments, our results can be easily compared to
the Lorentz-Dirac equation by writing it in the form
\begin{eqnarray}
\big( M \dot{v}^{\alpha} \big)_{\mathrm{LD}} &\simeq& - q
e_{a}^{\alpha} v_{b} F^{ab}_{\mathrm{ext}} + \frac{2}{3} q^{2}
\ddot{n}^{\alpha} +  o\Big(  \epsilon^{3} M T^{-1}
\max(\epsilon,\mathcal{E}) \Big)~, \label{LDe}
\\
\big( M n_{a} \dot{v}^{a} \big)_{\mathrm{LD}} &\simeq& -q n_{a}
e_{b}^{\beta} v_{\beta} F^{ab}_{\mathrm{ext}} + o\Big(  \epsilon^{3}
M T^{-1} \max(\epsilon,\mathcal{E}) \Big) ~. \label{LDn}
\end{eqnarray}

In order to avoid overcomplicating the discussion, we shall say that
our equations of motion reduce to the Lorentz-Dirac result if $M
\dot{v}^{\alpha}$ and $M n_{a} \dot{v}^{a}$ match (\ref{LDe}) and
(\ref{LDn}) up to terms of order $\epsilon^{2} M T^{-1}
\max(\epsilon^{2},\mathcal{E})$. Noting that $n_{a} \dot{v}^{a} = -
\dot{n}^{\alpha} v_{\alpha}$, (\ref{forcerad3}) shows that the the
`temporal component' of the body's acceleration always matches
(\ref{LDn}) to the required accuracy.

The same is not true for the spatial acceleration. This shouldn't be
too surprising, though, as the Lorentz-Dirac equation was never
intended to describe spinning particles. $S^{ab}$ should therefore
be set to zero (at least instantaneously) before any reasonable
comparison can be made. Even this isn't quite sufficient, though.
The situation can be remedied by assuming that
\begin{eqnarray}
|S| &<& \epsilon MD \max \big( \epsilon^{2}, \mathcal{E} \big) ~,
\label{SrestrictRad}
\\
\left| Q^{[a}{}_{c} F^{b]c}_{\mathrm{ext}} e_{a}^{\alpha}
e_{b}^{\beta}\right| &<& \epsilon^{2} M \max \big( \epsilon^{2},
\mathcal{E} \big)  ~, \label{dipolerestrictRad}
\end{eqnarray}
along with a similar restriction on $\left| Q^{[a}{}_{c}
F^{b]c}_{\mathrm{ext}} e_{a}^{\alpha} n_{b}\right|$. Then
\begin{eqnarray}
\left| v \right| &<&  \epsilon^{2} \max \big( \epsilon^{2},
\mathcal{E} \big) ~, \label{vRadLD}
\\
| \dot{M} | &<& \epsilon^{2} M T^{-1} \max \big( \epsilon^{2},
\mathcal{E} \big)  ~, \label{dotMRadLD}
\\
| \dot{S} | &<&  \epsilon^{2} M \max \big( \epsilon^{2}, \mathcal{E}
\big)  ~. \label{SboundRadLD}
\end{eqnarray}
If $\mathcal{E} \gtrsim \epsilon^{2}$, $\dot{M}$ vanishes up to the
maximum order that we can calculate it. The same is not necessarily
true for $\dot{S}^{\alpha\beta}$ and $v^{\alpha}$, although they
remain sufficiently small that (\ref{LDe}) can now be recovered to
the desired accuracy. There therefore exists a regime in which the
equations of motion derived here reduce to the Lorentz-Dirac
equation. Considering $z^{a}$ to be the only observable (as would be
reasonable for an extremely small particle), this completely
recovers the usual point particle result.

While this conclusion is not particularly surprising, it is
interesting to note how restrictive the required assumptions are.
(\ref{dipolerestrictRad}) is particularly difficult to satisfy. For
example, when the force is approximately given by the Lorentz
(monopole) expression, external field magnitudes up to $\sim M
\big(|q|T\big)^{-1}$ are allowed. In fields this large,
(\ref{dipolerestrictRad}) implies that the magnitudes of the dipole
moment must be less than $\sim \epsilon \mathcal{E} q D$ (when
$\mathcal{E} \gtrsim \epsilon^{2}$). This is an extremely limiting
condition in cases where the self-energy is small. Given
(\ref{dipole}), there do not appear to be any rigid ($\dot{\varphi}
= \dot{H}^{\alpha} =0$) or nearly rigid charge distributions that
could satisfy it. One either needs to choose a very special class of
charges, or considerably restrict the maximum allowable field
strength.

Before continuing, it should first be mentioned that our definitions
of the charge and three-current densities are slightly unusual.
These quantities would usually be defined with respect to an
orthonormal tetrad adapted to $z'^{a}\big(s(\bar{s})\big)$. We have
instead defined these quantities in terms of a tetrad with temporal
component $n^{a}$. Translating from one of these definitions to the
other would only introduce fractional changes of order
$\epsilon^{2}$, which are usually irrelevant at our level of
approximation.

\subsubsection{Retarded Self-Forces}
\label{RetardedSect}

It is generally accepted that detectors placed outside of $W$ will measure $F^{ab}_{\mathrm{ext}}+F^{ab}_{\mathrm{self}(-)}$ rather than $F^{ab}_{\mathrm{ext}}+F^{ab}_{\mathrm{self}(R)}$. But infinitesimal elements of an extended body cannot `know' that they are part of a larger whole, so they must couple to this same field. It is therefore reasonable to consider only the retarded self-field `physical.' Recalling that this is equal to the sum of the radiative and singular self-fields, any relevance of $F^{ab}_{\mathrm{self}(R)}$ itself should be derived by showing the effects of the singular self-field are irrelevant in certain cases.

To this end, we now examine the (presumably) realistic case in which the particle interacts with its full retarded self-field. Essentially all of the
steps in the previous section carry over identically in this case, although most involve considerably more calculation. Starting with
(\ref{chargedensity}), (\ref{threecurrent}), and
(\ref{field3})-(\ref{f4}), the retarded field can be shown to be
(dropping the `$-$' subscript)
\begin{eqnarray}
F^{ab}_{\mathrm{self}}(x) &\simeq& 2 \int \! \mathrm{d}^{3} r' \,
\frac{1}{R^{3}} \, \Bigg\{ \rho \left[ - (r-r')^{\alpha}
e_{\alpha}^{[a} n^{b]} \left(1+ \frac{1}{2} \dot{n}^{\beta}
(r-r')_{\beta} - \frac{1}{8} (\dot{n}^{\beta}(r-r')_{\beta})^{2}
+\frac{1}{8} R^{2} |\dot{n}|^{2} \right. \right.
\nonumber
\\
&& \left. {} + \frac{1}{2} \dot{n}^{\beta} \dot{n}^{\gamma}
r_{\beta} (r-r')_{\gamma}  \right) + \frac{1}{2} R^{2}
(r-r')^{\alpha} e_{\alpha}^{[a} \ddot{n}^{b]} - \frac{1}{2} R^{2}
n^{[a} \dot{n}^{b]} \left( 1+ \frac{1}{2} \dot{n}^{\beta}
(3r-r')_{\beta} \right) \nonumber
\\
&& \left. {} - \frac{2}{3} R^{3} \ddot{n}^{[a} n^{b]} \right] -
\Big( H^{\beta} + v^{\sigma} \partial'_{\sigma} \left( r'^{\beta}
\varphi \right) \Big) \left[ (r-r')^{\alpha} e_{\alpha}^{[a}
e^{b]}_{\beta} \left( 1 + \frac{1}{2} \dot{n}^{\gamma}
(r+r')_{\gamma} + \frac{3}{8} \Big(\dot{n}^{\gamma}
(r+r')_{\gamma}\Big)^{2} \right. \right. \nonumber
\\
&& \left. {} + \frac{1}{8} R^{2} |\dot{n}|^{2} - \frac{1}{2}
\dot{n}^{\gamma} \dot{n}^{\lambda} r_{\gamma} r'_{\lambda} \right) +
n^{[a} e^{b]}_{\beta} R^{2} \left( \frac{1}{2} \ddot{n}^{\gamma}
(r+r')_{\gamma} + \frac{1}{3} R |\dot{n}|^{2} - R^{-2} v^{\gamma}
(r-r')_{\gamma} \right) \nonumber
\\
&& \left. {} + \frac{1}{2} R^{2} \dot{n}_{\beta} (r-r')^{\alpha}
e_{\alpha}^{[a} \dot{n}^{b]} + \frac{1}{2} R^{2} \ddot{n}_{\beta}
(r-r')^{\alpha} e^{[a}_{\alpha} n^{b]} + \frac{1}{2} R^{2}
\dot{n}^{[a} e^{b]}_{\beta} \left( 1+ \frac{1}{2}
\dot{n}^{\gamma}(3r+r')_{\gamma} \right) \right. \nonumber
\\
&& \left. {} - \frac{1}{3} R^{3} \ddot{n}^{[a} e^{b]}_{\beta}
\right] + \dot{\varphi} (r-r')^{\alpha} e_{\alpha}^{[a}
e_{\beta}^{b]} r'^{\beta} - \dot{H}^{\beta} n^{[a} e^{b]}_{\beta}
R^{2} \Bigg\} ~. \label{field4}
\end{eqnarray}

The self-force is now obtained by inserting this into
(\ref{forcedefine}), which results in an expression of the form
\begin{eqnarray}
\Psi^{a}_{\mathrm{self}}(s) &\simeq& - \int \! \mathrm{d}^{3} r \,
\mathrm{d}^{3} r' \, \mathcal{F}^{a}(r,r',s) ~,
\\
&\simeq& - \frac{1}{2} \int \! \mathrm{d}^{3} r \, \mathrm{d}^{3} r'
\, \Big( \mathcal{F}^{a}(r,r',s) + \mathcal{F}^{a}(r',r,s) \Big) ~,
\label{Newton3}
\end{eqnarray}
where $\mathcal{F}^{a}(r,r',s)$ represents the force density exerted
by a charge element at $r$ on a charge element at $r'$. If Newton's
third law were correct, $\mathcal{F}^{a}(r,r',s) = -
\mathcal{F}^{a}(r',r,s)$ (so $\Psi^{a}_{\mathrm{self}}$ would
vanish). Of course, this is does not quite hold for the
electromagnetic field (or any other fully observable field), so
there is a nonzero self-force. (\ref{Newton3}) therefore gives a
precise form to the intuitive idea that self-forces measure the
degree of failure of Newton's third law.

After removing the components of $\mathcal{F}^{a}$ which reverse
sign under interchange of $r$ and $r'$, it can be shown that
\begin{eqnarray}
M \dot{n}^{\alpha} &\simeq& -\Psi^{\alpha}_{\mathrm{ext}} +
\frac{2}{3} q^{2} \ddot{n}^{\alpha} - M_{\mathrm{em}}
\dot{n}^{\alpha} - \int \! \mathrm{d}^{3}r \, \mathrm{d}^{3} r' \,
\frac{1}{R^{3}} \, \Bigg\{ \rho(r,s) H^{\beta}(r',s) R^{2} \left[
\delta^{\alpha}_{\beta} \bigg( \frac{1}{2} \ddot{n}^{\gamma}
(r+r')_{\gamma} \right. \nonumber
\\
&& \left. {} - R^{-2} v^{\gamma} (r-r')_{\gamma} \bigg) -
(r-r')^{\alpha} \ddot{n}_{\beta} \right]  + \rho(r,s)
\dot{H}^{\beta}(r',s) R^{2} + v^{\gamma} \partial_{\gamma} \Big(
r_{\beta} \varphi(r,s) \Big) H^{\alpha}(r',s) (r-r')^{\beta}
\nonumber
\\
&& {} -\dot{\varphi}(r,s) H^{\gamma}(r',s) (r-r')^{\alpha}
r_{\gamma} - \left( \varphi(r,s) - \frac{q}{4\pi |r|^{3}} \right)
\left[ \frac{1}{2} H^{\gamma}(r',s) \bigg( R^{2} \ddot{n}^{\alpha}
r_{\gamma} - R^{2} \delta^{\alpha}_{\gamma} \ddot{n}^{\beta}
r_{\beta} \right. \nonumber
\\
&& \left. {} + \Big( (r-r')^{\alpha} r_{\gamma} -
\delta^{\alpha}_{\gamma} (r-r')^{\beta} r_{\beta} \Big)
\ddot{n}^{\sigma} (r-r')_{\sigma} \bigg) + \dot{H}^{\gamma}(r',s)
\left( (r-r')^{\alpha} r_{\gamma} - \delta^{\alpha}_{\gamma}
(r-r')^{\beta} r_{\beta} \right) \right] \Bigg\}
\nonumber
\\
&& {} + o \big( \epsilon^{2} \mathcal{E} M T^{-1} \big) ~,
\label{fullforce}
\end{eqnarray}
where it was useful to define an `electromagnetic mass'
\begin{eqnarray}
M_{\mathrm{em}} &:=& \int \! \mathrm{d}^{3} r \, \mathrm{d}^{3} r'
\, \frac{1}{R^{3}} \, \Bigg\{  \frac{1}{2} R^{2} \bigg[  \rho(r,s)
\rho(r',s) - H_{\beta}(r,s) H^{\beta}(r',s) \Big( 1+
\dot{n}^{\gamma} (r+r')_{\gamma} \Big) \bigg] \nonumber
\\
&& {} + \left( \varphi(r,s) - \frac{q}{4\pi |r|^{3}} \right)
\rho(r',s) \left[ (r-r')^{\beta} r_{\beta} \left( 1 + \frac{1}{2}
\dot{n}^{\gamma} (r-r')_{\gamma} \right) - \frac{1}{2} R^{2}
\dot{n}^{\beta} r_{\beta} \right] \Bigg\} ~. \label{dM}
\end{eqnarray}

The form of (\ref{fullforce}) suggests an effective inertial mass $m
:= M + M_{\mathrm{em}}$. Although $M$ will very rarely remain
constant, $m$ is often conserved, or at least varies slowly. To see
this, substitute (\ref{field4}) into (\ref{forcedefine}) and use
(\ref{massevolve}) to show that
\begin{eqnarray}
\dot{M} &\simeq& -n_{a} \Psi^{a}_{\mathrm{ext}} - \int \!
\mathrm{d}^{3} r \, \mathrm{d}^{3} r' \, \frac{1}{R^{3}} \, \Bigg\{
\left( \varphi(r,s) -\frac{q}{4\pi |r|^{3}} \right) \bigg[
\dot{\rho}(r',s) (r-r')^{\beta} r_{\beta} - \frac{1}{2} \rho(r',s)
\Big( R^{2} \ddot{n}^{\beta} r_{\beta} \nonumber
\\
&& {} -\ddot{n}^{\alpha} (r-r')_{\alpha} (r-r')^{\beta} r_{\beta}
\Big) -  H^{\gamma}(r',s) \left( \frac{1}{2} R^{2} \Big(
\dot{n}_{\gamma} \dot{n}^{\beta} r_{\beta} + r_{\gamma}
|\dot{n}|^{2} \Big)  - (r-r')^{\alpha} \Big( \dot{n}_{\alpha}
r_{\gamma} \right. \nonumber
\\
&& {} - r_{\alpha} \dot{n}_{\gamma} \Big) \Big(1+ \frac{1}{2}
\dot{n}^{\sigma} (r+r')_{\sigma} \Big) \bigg) \bigg]  -H_{\beta}(r)
\dot{H}^{\beta}(r') R^{2} - \frac{1}{2} H_{\beta}(r) H^{\beta}(r')
R^{2} \ddot{n}^{\gamma}(r+r')_{\gamma} \nonumber
\\
&& {}  - \rho(r',s) \bigg[ R^{2} H_{\beta}(r,s) \dot{n}^{\beta}
\left( 1+ \frac{1}{2} \dot{n}^{\alpha} (3 r-r')_{\alpha} \right)  -
v^{\gamma} \partial_{\gamma} \Big(r_{\beta} \varphi(r,s) \Big)
(r-r')^{\beta} \bigg] \Bigg\} + o \big( \epsilon^{2} \mathcal{E} M
T^{-1} \big) ~. \label{dotM}
\end{eqnarray}
Combining this with (\ref{dM}) and the definition of $m$,
\begin{eqnarray}
\dot{m} &\simeq & -n_{a} \Psi^{a}_{\mathrm{ext}} +  \int \!
\mathrm{d}^{3} r \, \mathrm{d}^{3} r' \, \frac{1}{R^{3}} \, \Bigg\{
\left( \varphi(r,s) - \frac{q}{4\pi |r|^{3}} \right)
H^{\gamma}(r',s) \bigg[ \Big( \dot{n}_{\gamma} (r-r')^{\alpha}
r_{\alpha} - r_{\gamma} \dot{n}^{\alpha} (r-r')_{\alpha} \Big)
\nonumber
\\
&& {} \times \Big( 1  + \frac{1}{2} \dot{n}^{\lambda}
(r+r')_{\lambda} \Big) + \frac{1}{2} R^{2} \Big( |\dot{n}|^{2}
r_{\gamma} + \dot{n}^{\beta} r_{\beta} \dot{n}^{\gamma} \Big)
\bigg] + \rho(r',s) \bigg[ R^{2} H_{\beta}(r,s) \dot{n}^{\beta}
\left( 1 + \frac{1}{2} \dot{n}^{\alpha} (3r-r')_{\alpha} \right)
\nonumber
\\
&& {} - v^{\gamma} \partial_{\gamma} \Big( r_{\beta} \varphi(r,s)
\Big) (r-r')^{\beta} \bigg] \Bigg\} + o\big( \epsilon^{2}
\mathcal{E} M/T \big) ~, \label{dotm}
\end{eqnarray}
which is considerably simpler than (\ref{dotM}).

Although we now have all of the results necessary to compute each
tetrad component of $\Psi^{[ab]}_{\mathrm{self}}$ to second order,
the resulting expressions are extremely lengthy (and correspondingly
difficult to interpret). It was also seen when examining the
radiative self-fields that the second-order terms in the self-torque
were not necessary to reach the point particle limit. For both of
these reasons, we only compute it here to first order. Combining
(\ref{torquedefine}) and (\ref{field4}),
\begin{eqnarray}
\Psi^{[\alpha \beta]}_{\mathrm{self}} & \simeq & \int \!
\mathrm{d}^{3} r \, \mathrm{d}^{3} r' \, \frac{\dot{n}^{[\alpha}
r^{\beta]}}{R} \, \Bigg[ \frac{1}{2} \bigg( \rho(r,s) \rho(r',s) +
H_{\gamma}(r,s) H^{\gamma}(r',s) \bigg) + \left( \varphi(r,s) -
\frac{q}{4\pi |r|^{3}} \right) \rho(r',s) \Bigg] \nonumber
\\
&& {} + o \big( \epsilon^{2} \mathcal{E} M \big) ~, \label{torque}
\\
n_{a} e_{b}^{\beta} \Psi^{[ab]}_{\mathrm{self}} &\simeq& \frac{1}{2}
\int \! \mathrm{d}^{3}r \, \mathrm{d}^{3} r' \, \frac{1}{R^{3}} \,
\Bigg\{ \rho(r,s) H_{\alpha}(r',s) r'^{\beta} \left[ (r-r')^{\alpha}
\Big( 1 - \frac{1}{2} \dot{n}^{\gamma} (r-r')_{\gamma} \Big) +
\frac{1}{2} R^{2} \dot{n}^{\alpha} \right] \nonumber
\\
&& {} - \left( \varphi(r,s) - \frac{q}{4\pi |r|^{3}} \right) \bigg[
R^{2} r_{\gamma} \dot{n}^{[\beta} H^{\gamma]}(r',s) + 2 r_{\gamma}
(r-r')^{[\beta} H^{\gamma]}(r',s) \Big( 1 + \frac{1}{2}
\dot{n}^{\gamma} (r+r')_{\gamma} \Big) \nonumber
\\
&& {} - r_{\alpha} H^{\alpha}(r',s) \Big( r^{\beta} \dot{n}^{\gamma}
r'_{\gamma} - r'^{\beta} \dot{n}^{\gamma} r_{\gamma} \Big) +
(r-r')^{\alpha} r_{\alpha} \Big( H^{\beta}(r',s) \dot{n}^{\gamma}
r_{\gamma} - r^{\beta} \dot{n}_{\gamma} H^{\gamma}(r',s) \Big)
\bigg] \Bigg\} \nonumber
\\
&& {} + o \big( \epsilon^{2} \mathcal{E} M \big) ~. \label{ntorque}
\end{eqnarray}

\section{Special Cases}
\label{SpecCaseSect}

\subsection{Current-Dominated Particles}

The full equations of motion derived above are obviously quite
complicated,, so it is useful to specialize the discussion somewhat.
First assume that the particle's charge density is sufficiently
small that it can be entirely dropped without losing any accuracy.
This requires that $\varphi < \epsilon^{2} H$. Then (\ref{torque})
becomes equivalent to
\begin{equation}
\dot{S}^{\alpha\beta} \simeq - 2 \Psi^{[\alpha
\beta]}_{\mathrm{ext}} + \int \! \mathrm{d}^{3}r \, \mathrm{d}^{3}r'
\, \frac{r^{[\alpha} \dot{n}^{\beta]}}{R} \, H_{\gamma}(r,s)
H^{\gamma}(r',s) + o \big( \epsilon^{2} \mathcal{E} M \big) ~.
\label{SdotCurrent1}
\end{equation}
This can be quite large. In order to make sure that it satisfies
(\ref{torquerestrict}), let
\begin{equation}
\mathcal{E} < \epsilon ~. \label{selfenergyrestrict2}
\end{equation}

Now, $M_{\mathrm{em}}  \lesssim \epsilon M$, so the remaining
equations of motion can be derived from  (\ref{fullforce}),
(\ref{dotm}), and (\ref{ntorque}):
\begin{eqnarray}
\dot{m} &\simeq& - n_{a} \Psi^{a}_{\mathrm{ext}} + o \big(
\epsilon^{3} m T^{-1} \big) ~,
\\
m \dot{n}^{\alpha} &\simeq& - \Psi^{\alpha}_{\mathrm{ext}} + o\big(
\epsilon^{3} m T^{-1} \big) ~, \label{ndotevolveCurrent1}
\\
m v^{\alpha} &\simeq& S^{\alpha \beta} \dot{n}_{\beta} - 2
\Psi^{[ab]}_{\mathrm{ext}} e_{a}^{\alpha} n_{b} + o \big(
\epsilon^{3} m \big) ~. \label{CMevolveCurrent1}
\end{eqnarray}

An expression for $\dot{v}^{\alpha}$ is obtained by differentiating
(\ref{CMevolve}), which gives
\begin{eqnarray}
m \dot{v}^{\alpha} &\simeq& - \Psi^{\alpha}_{\mathrm{ext}} + n_{a}
\Psi^{a}_{\mathrm{ext}} v^{\alpha} + S^{\alpha \beta}
\ddot{n}_{\beta} - 4 \Psi^{[\alpha \beta]}_{\mathrm{ext}}
\dot{n}_{\beta} - 2 \dot{\Psi}^{[ab]}_{\mathrm{ext}} e_{a}^{\alpha}
n_{b}  \nonumber
\\
&& {} + 2 \int \! \mathrm{d}^{3} r \, \mathrm{d}^{3} r' \,
\frac{r^{[\alpha} \dot{n}^{\beta]} \dot{n}_{\beta} }{R} \,
H_{\gamma} (r,s) H^{\gamma} (r',s) + o \big( \epsilon^{3} m T^{-1}
\big) ~.
\end{eqnarray}
The presence of a (potentially) significant self-torque here is
completely different than the situation that arose when considering
only the radiative component of the self-field. Even adopting
conditions (\ref{lambdarestrict}), (\ref{SrestrictRad}), and
(\ref{dipolerestrictRad}) would not generically recover the
Lorentz-Dirac equation.

Instead, we find (assuming $\mathcal{E} \gtrsim \epsilon^{2}$) that
(\ref{vRadLD}) and (\ref{dotMRadLD}) (with $\dot{M} \rightarrow
\dot{m}$) would remain correct, although (\ref{SboundRadLD}) is
weakened to $|\dot{S}| < \epsilon^{2} m$. In this case, the
center-of-mass line is governed by
\begin{equation}
m \dot{v}^{\alpha} \simeq - q e_{a}^{\alpha} v_{b}
F^{ab}_{\mathrm{ext}} + \frac{2}{3} q^{2} \ddot{n}^{\alpha} + 2
M_{\mathrm{em}} D_{\mathrm{H}}^{[\alpha} \dot{v}^{\beta]}
\dot{v}_{\beta} + o \big(\epsilon^{3} m T^{-1} \big) ~,
\end{equation}
where
\begin{equation}
D^{\alpha}_{\mathrm{H}}(s)  :=  \frac{1}{M_{\mathrm{em}}} \int \!
\mathrm{d}^{3} r \, \mathrm{d}^{3} r' \left( \frac{r^{\alpha} }{R}
\right)  H_{\gamma} (r,s) H^{\gamma} (r',s) ~.
\end{equation}
Given (\ref{dM}), $D^{\alpha}_{\mathrm{H}}$ appears to be related to
the shift in the position of the effective center-of-mass due to the
electromagnetic self-energy.

The force that it generates clearly becomes irrelevant when
$M_{\mathrm{em}} D^{[\alpha}_{\mathrm{H}} \dot{v}^{\beta]}
\dot{v}_{\beta} \lesssim \epsilon^{3} m/T$, which occurs when
$D^{\alpha}_{\mathrm{H}}$ (nearly) coincides with
$\dot{v}^{\alpha}$, or more generically if
\begin{equation}
\left| D_{\mathrm{H}} \right| \lesssim \epsilon D ~.
\label{Hrestrict1}
\end{equation}
This happens, for example, in cases where $H^{\gamma}(r,s) \simeq
\pm H^{\gamma} (-r,s)$. (\ref{Hrestrict1}) can therefore be thought
of as a restriction on the allowed asymmetry in the particle's
current structure about the center-of-mass. Given (\ref{dipole}), a
similar intuitive interpretation can also be applied to
(\ref{dipolerestrictRad}), so there is some overlap between these
two conditions. It does not appear that either one strictly implies
the other, however.

Also note that if (\ref{Hrestrict1}) holds,
(\ref{selfenergyrestrict2}) is no longer required. The case
$\mathcal{E} \sim 1$ is then quite interesting if the spin and/or
dipole moment are non-negligible. When this happens, the left-hand
side of (\ref{CMevolveCurrent1}) needs to be replaced by
$Mv^{\alpha}$, which means that $M$ and $m$ must both be kept track
of. The expression for $\dot{M}$ is not simple, so this is a
considerable complication.

\subsection{Charge-Dominated Particles}

Treating the opposite case, we now consider particles with very
small internal currents. In particular, let $H < \epsilon^{2}
\varphi$. Given (\ref{HBoundary}), this is about as small as $H$
could possibly be without fine-tuning. (\ref{threecurrent}) and
(\ref{phisize}) show that the internal currents which arise due to
charges rearranging themselves (via elasticity, rotation, etc.) are
bounded by this same amount. So we actually have that $|j| \lesssim
\epsilon^{2} \varphi$.

(\ref{torque}) now reduces to
\begin{equation}
\dot{S}^{\alpha \beta} \simeq - 2 \Psi^{[\alpha
\beta]}_{\mathrm{ext}} + \int \! \mathrm{d}^{3} r \, \mathrm{d}^{3}
r' \, \frac{ r^{[\alpha} \dot{n}^{\beta]} }{R} \, \rho(r',s) \left[
\rho(r,s) + 2 \left( \varphi(r,s) - \frac{q}{4\pi |r|^{3}} \right)
\right] + o \big( \epsilon^{2} \mathcal{E} m \big) ~.
\end{equation}
Once again, this expression can easily violate
(\ref{torquerestrict}) unless $\mathcal{E} < \epsilon$. Assuming
this,
\begin{eqnarray}
\dot{m} &\simeq& -n_{a} \Psi^{a}_{\mathrm{ext}} + \int \!
\mathrm{d}^{3}r \, \mathrm{d}^{3}r' \, \frac{1}{R} v^{\alpha}
\partial_{\alpha} \rho(r,s) \rho(r',s) + o \left(
\epsilon^{3} m T^{-1} \right) ~,
\\
m \dot{n}^{\alpha} &\simeq& - \Psi^{\alpha}_{\mathrm{ext}} +
\frac{2}{3} q^{2} \ddot{n}^{\alpha} + o \left( \epsilon^{3} m T^{-1}
\right) ~,
\\
m v^{\alpha} &\simeq& S^{\alpha \beta} \dot{n}_{\beta} - 2
\Psi^{[ab]}_{\mathrm{ext}} e_{a}^{\alpha} n_{b} + o\left(
\epsilon^{3} m \right) ~.
\end{eqnarray}

By analogy to the current-dominated case, we can now define
\begin{equation}
D_{\mathrm{\varphi}}^{\alpha} := \frac{1}{M_{\mathrm{em}}} \int \!
\mathrm{d}^{3} r \, \mathrm{d}^{3} r' \left( \frac{ r^{\alpha} }{R}
\right) \rho(r',s) \left[ \rho(r,s) + 2 \left( \varphi(r,s) -
\frac{q}{4\pi |r|^{3}} \right) \right] ~.
\end{equation}
Again, this looks like the center-of-electromagnetic mass, although
the interpretation isn't quite so direct as it was for
$D_{\mathrm{H}}^{\alpha}$.

Now let $\mathcal{E} \gtrsim \epsilon^{2}$, and adopt
(\ref{lambdarestrict}), (\ref{SrestrictRad}), and
(\ref{dipolerestrictRad}). As before, these conditions imply that
$|v| < \epsilon^{3}$, so $\dot{m} < \epsilon^{3} m T^{-1}$. The
Lorentz-Dirac equation is recovered when $M_{\mathrm{em}}
D_{\mathrm{\varphi}}^{[\alpha} \dot{v}^{\beta]} \dot{v}_{\beta}
\lesssim \epsilon^{3} m T^{-1}$. If the acceleration is not
restricted to be in a specific direction, then it is convenient to
satisfy this condition by requiring
\begin{equation}
\left| D_{\mathrm{\varphi}} \right| \lesssim \epsilon D ~.
\label{Phirestrict1}
\end{equation}
This clearly holds when  (for example) $\varphi(r,s) \simeq \pm
\varphi(-r,s)$, so $|D_{\mathrm{\varphi}}|/D$ can be considered a
measure of the charge's internal symmetry.

The two examples discussed here are by the far the simplest,
although other ones may also be of interest. For example, the
condition $|j| \lesssim \epsilon \varphi$ could have replaced $|j|
\lesssim \epsilon^2 \varphi$ in this section. This would allow for
some charge-current coupling without introducing undue complexity.
Similarly, one might be interested in the case where $\varphi
\lesssim \epsilon H$, or the general scenario where $\varphi$ and
$H^{\alpha}$ are symmetric (or antisymmetric) about $r=0$.

\section{Conclusions}

\subsection{Results}

Starting from the fundamental equations
(\ref{maxwell})-(\ref{stresscons}), we have now derived equations of
motion describing a very wide variety of classical extended charges
in flat spacetime. One of our main motivations has been to
investigate the validity of the commonly-held notion that `small'
charges can often be treated as though they were perfectly
pointlike. Because of this, the precise implications of each
approximation have been emphasized, and all of our results have been
kept as general as (reasonably) possible.

To review, charge distributions were considered where all
significant length scales remained of order $D(s)$ within each time
slice $\Sigma(s)$. An acceleration timescale $T\gg D$ was then
defined to place lower bounds on $|\dot{n}|^{-1}$ and
$|\ddot{n}|^{-1/2}$. This was assumed to relate to the timescales of
the body's internal motions as well. Specifically, the
$s$-derivatives of the `charge potential' $\varphi$ were required to
satisfy (\ref{phisize}) and (\ref{phisize2}). Similar restrictions
were placed on the derivatives of $H^{\alpha}$ as well. The last
major restriction required that the spin and torque play a
relatively small but non-negligible role in the system's behavior.
This was formalized by assuming that $|n-v|=|v|$ remained less than
about $\epsilon^{2}=(D/T)^{2}  \ll 1$. All of these assumptions are
kinematic, and cannot automatically be assumed to hold without any
regard for the sizes of the external fields, spin, and so on.
Self-consistency is preserved by requiring the angular momentum,
force, and torque obey the bounds discussed in Sec.
\ref{Approximations}.

The restriction given on $S^{ab}$ essentially states that the body's
rotational period can't be less than $T$. The relations satisfied by
the force and torque are not quite as simple, though. Both of these
quantities depend on $J^{a}$, so any bounds placed on them must
affect the class of allowable current distributions. While this can
be taken into account in different ways, we have chosen to impose
restrictions on the dipole moment and the `relative self-energy'
$D_{\mathrm{em}}/D=\mathcal{E}$. If the dominant force is just $-q
F^{ab}_{\mathrm{ext}} v_{b}$, $Q^{ab}$ must satisfy
(\ref{dipolerestrict}).

The bound on $\mathcal{E}$ is more complicated to summarize, as it
is sensitive to the specific case under consideration. The
calculation using the radiative self-field is the least restrictive;
allowing all $\mathcal{E} \lesssim 1$. One could technically let the
electromagnetic radius exceed the physical one in this case,
although we take the point of view that this would be too
unphysical. Regardless, taking into account the full retarded
self-field shows that there exists a significant class of charge
distributions where the self-torque will become too large whenever
$\mathcal{E} \gg \epsilon^{2}$. In scenarios where the
charge-current coupling is negligible, the relative self-energy can
be of order $\epsilon$. Still, there are certain special cases such
as spherical symmetry where the restriction on $\mathcal{E}$ can be
completely relaxed.

Despite appearances, these assumptions are really no different than
the standard slow-motion approximation. Using it, the radiative and
retarded self-fields in a neighborhood of $W$ were shown to be
approximated by (\ref{radfield}) and (\ref{field4}) respectively.
These expressions are quite general, and although the retarded field
is written using the peculiar constructions of Dixon's theory, it
can easily be translated into a more conventional notation by using
(\ref{chargedensity}) and (\ref{threecurrent}).
(\ref{field3})-(\ref{f4}) also provide a convenient starting point
for this.

These expressions for the fields were then combined with the exact
forms of Dixon's equations of motion to yield the results contained
in Sec. \ref{ForceSect}. All of the approximations adopted here were
therefore applied only to compute the fields. In the (unrealistic)
case that the body was assumed to couple only to the radiative
component of its self-field, the center-of-mass line was found to
evolve according to (\ref{dotmrad1})-(\ref{forcerad2}). The external
forces and torques that appear in these expressions are given
exactly by (\ref{forcedefine}) and (\ref{torquedefine}), or more
intuitively by (\ref{forcemult}) and (\ref{torquemult}).

In order to investigate the Lorentz-Dirac limit, it was first
necessary to require the particle's radius to be sufficiently small
that the dipole and higher contributions to the external force could
be considered negligible without having to throw away the
Lorentz-Dirac component of the self-force. This idea was summarized
in the condition (\ref{lambdarestrict}), which transformed the
equations of motion in the radiative case into
(\ref{dotmrad2})-(\ref{vrad1}). Recovering the Lorentz-Dirac
equation was then seen to require bounding the angular momentum and
dipole moment by (\ref{SrestrictRad}) and (\ref{dipolerestrictRad})
respectively.

Since this limit is meant to describe a non-spinning particle, the
bound on the angular momentum is hardly surprising.
(\ref{dipolerestrictRad}) is rather different, though. It shows that
even when ignoring the `singular' portion of the self-field, the
Lorentz-Dirac equation does not apply to all small charge
distributions. If a problem were to involve very large (external)
field strengths and small self-energies, the class of charges which
move like a point particle would in fact become extremely small.
Oddly enough, this problem disappears for large self-energies. In
these cases, the restriction on the dipole moment is no stronger
than was required for self-consistency of the initial slow-motion
assumptions.

This issue is made considerably more complex when the body is
allowed to interact with its full retarded self-field. Without
placing any restrictions on the spin or dipole moment,
$\dot{n}^{\alpha}$ was found to be given by (\ref{fullforce}). This
involves an effective mass $m=M+M_{\mathrm{em}}$, which involves an
electromagnetic contribution given by (\ref{dM}). It does not
generally remain constant, but rather evolves according to
(\ref{dotm}). There are also self-torques given by (\ref{torque})
and (\ref{ntorque}). Combining these expressions with
(\ref{tdefine}) and (\ref{CMevolve}) recovers the full equations of
motion. These are our central result. They provide a considerable
generalization of the Lorentz-Dirac equation.

They are also much more complicated than the equations derived using
only the radiative portion of the self-field. The majority of this
complexity arises from interactions between $\varphi$ and
$H^{\alpha}$, which -- given (\ref{chargedensity}) and
(\ref{threecurrent}) -- can be viewed as charge-current couplings in
the center-of-mass frame. Predictably then, our results are
considerably simplified when only one of these quantities is
significant. $\varphi$ will (generically) drop out of the equations
of motion only if it is of order $\epsilon^{2} H$ or less. The
reverse is also true. Both of these cases were discussed in detail
in Sec. \ref{SpecCaseSect}. It should be emphasized that these
restrictions on the relative magnitudes of the charge and current
densities are exceptionally strong. Simply saying that one of these
quantities is much larger than the other without any further
qualification is not sufficient to remove the coupling terms.

Even in these cases, though, the previous restrictions on $\lambda$,
$S^{ab}$, and $Q^{ab}$ must be supplemented by either
(\ref{Hrestrict1}) or (\ref{Phirestrict1}) in order to recover the
Lorentz-Dirac equation. Intuitively, these relations ensure that the
`center-of-electromagnetic mass' is not too far from $r=0$. This is
roughly what was already required by (\ref{dipolerestrictRad}),
although the two conditions are not quite the same.

Note that the assumptions we have given to be able to derive the
Lorentz-Dirac equation are not exhaustive. They were the most
obvious choices obtained by examining our general equations of
motion, but are slightly more restrictive than necessary. Still,
they seem to be reasonably effective for most systems that are not
finely-tuned.

\subsection{Discussion}
\label{discussion}

In summary, it was shown that the Lorentz-Dirac limit is rather
delicate, and that the radiative self-field is rarely an adequate
replacement for the retarded one. It is now interesting to speculate
how these results might generalize in curved spacetime. One might be
interested in the motion of a charged particle moving in a
background spacetime, an uncharged body allowed to generate its own
gravitational field, or even the general case of a massive charge.
With the exception of this last possibility (which doesn't
necessarily follow from the two simpler calculations), such systems
have been considered in the past by a number of authors
\cite{NodvikOrig, MST, DewittBrehme, Hobbs, SFReview1, QuinnWald,
PN1}. All of these calculations have either applied a point particle
ansatz, or considered only very special (though usually
astrophysically motivated) classes of extended bodies. Both such
methods have agreed with each other, although it is clear that
\textit{some} extended bodies must exist which do not move like
point particles.

Finding the scope of the existing MiSaTaQuWa equation
\cite{SFReview1} -- as well as its generalization -- would likely
require an analysis similar to the one given here. The methods used
in this paper have in fact been chosen specifically for their
ability to be easily applied to fully dynamic spacetimes (except for those used to expand the self-field). To see
this, it should first be mentioned that all of the advantages of
Dixon's formalism are retained without approximation in full general
relativity \cite{Dix74,Dix79}.

This provides two essential results. The first of these is a natural
notion of a center-of-mass line. This definition has been proven to
satisfy nearly all of the intuitive notions that one might expect of
such an object \cite{Schatt1}, so it can be considered a reasonable
measure of a particle's `average' position. Just as importantly, it
is exactly determined by a finite number of ordinary differential
equations \cite{Dix79, Ehl77}. The analogous equations here were
(\ref{massevolve})-(\ref{CMevolve}), and (\ref{tdefine}). These
expressions are only slightly complicated by the transition to
curved spacetime.

The second contribution of Dixon's formalism is the set of
stress-energy moments itself. These are all contained in the
`stress-energy skeleton' $\hat{T}^{ab}(r,s)$. Here $r$ is a
coordinate in the tangent space of $z(s)$, which is an essential
point that was not obvious in flat spacetime. Importantly, the
constraints on $\hat{T}^{ab}$ implied by the generalized form of
(\ref{stresscons}) are exactly the same in all spacetimes. This
obviously includes the flat case summarized in the appendix. Just as
(\ref{hatJ}) gave the general form for any current skeleton
$\hat{J}^{a}$ satisfying the constraints, one can also find all
possible forms of $\hat{T}^{ab}$ \cite{Dix74}.

As with current skeleton, not all of these possibilities are
physically reasonable. Some will be singular, and others will have
supports extending to spatial infinity. We conjecture that such
cases can be removed exactly as they were in the appendix for the
current moments. This would leave a set of `reduced moment
potentials' that could be arbitrarily generated from some simple
recipe. An automatic byproduct of this reduction process would be a
relatively straightforward method of generating $T^{ab}$ from the
potentials. The analogous results in this paper were that the
current skeleton was determined by the freely-specifiable functions
$\varphi$ and $\bar{H}^{\alpha \beta}$. These in turn generated
$J^{a}$ via (\ref{chargedensity}), (\ref{threecurrent}), and
(\ref{divH}).

In the end, such reduced potentials for the stress-energy moments
would serve (in combination with solutions of the ODE's for the
linear and angular momenta) as a nearly background-independent way
of specifying physically reasonable conserved stress-energy tensors.
Besides the intrinsic elegance of such a construction, it would also
solve several problems that did not arise in the present paper. The
most important of these is the location of the center-of-mass. At
any $s$, this is clearly determined by $T^{ab}$ on $\Sigma(s)$. The
definition is highly implicit, however, and it is almost always
impractical to apply in practice. Specifying the matter via
$\hat{T}^{ab}$ would avoid this problem. Constructing it in the
natural way would \textit{start} with the center-of-mass position.
It would also incorporate $p^{a}$ and $S^{ab}$ automatically. These
quantities obviously must solve certain differential equations, and
writing down a $T^{ab}$ on each time slice with the correct momenta
would be extremely difficult with any generality. These issues were
the main motivations for discussing the current moments in so much
detail in the appendix, as well as the use of $\varphi$ and
$H^{\alpha}$ throughout this paper instead of $J^{a}$. These choices
slightly obscure the results here, but would be essential in any
generalization.

With this formalism in place, one would then need to compute the
metric to find the body's motion. As in the electromagnetic case,
this is the most difficult step. Of course, Einstein's equation is
considerably more complicated than Maxwell's, so it may be
prohibitive to carry out the calculation by hand (except in special
cases). Regardless, the above procedure could be integrated into
existing numerical relativity codes. This would then allow one to
rigorously study spacetimes with nonsingular matter fields without
having to solve the conventional elasticity (or Navier-Stokes)
equations. The most convenient such systems in this formalism may
not represent the most astrophysically interesting types of matter,
although there is no shortage of important problems in numerical
relativity where the details of the matter distributions are not a
primary concern.

Whatever the results of such inquiries, our results in
electromagnetism can be used to speculate how point particle methods
might break down in gravitational self-force problems. For this, it
is useful to give an intuitive explanation for such failures in
Maxwell's theory. The simplest of these derives from the fact that
the center-of-mass does not generally correspond to anything that
could be called a `center-of-charge.' In these cases,
electromagnetic self-forces effectively act through a lever arm.
This induces a torque, which in turn affects the particle's overall
motion. The situation is made particularly complicated when the
self-force is broken up into several pieces (as is natural). Each
such piece tends to act through a different point, and there's no
particular reason that any of them should coincide with the
center-of-mass (although the initial restriction
(\ref{dipolerestrict}) on the magnitude of the dipole moment does
help).

In contrast, one might expect that gravitational self-forces would
always act through the center-of-mass; effectively removing this
type of effect. This isn't necessarily true, though. The
center-of-mass definition only requires that $t^{abc}n_{b}n_{c}=0$,
which does not necessarily mean that $t^{abc}$ completely vanishes.
The portion of the self-force acting as an effective mass could also
produce a significant torque.  (\ref{stressdipole}) shows, however,
that $t^{abc}$ does vanish whenever $S^{ab} \rightarrow 0$. This is
the only case in which the MiSaTaQuWa equation can be reasonably
expected to hold, so imposing it at the start should remove any
problems. Since the dipole moment here is entirely dependent on
$S^{ab}$ (unlike in the electromagnetic case) it should be
relatively easy to arrange for an initially-vanishing angular
momentum to remain small. These sorts of effects might therefore be
expected to be less troublesome in gravity than in electromagnetism,
although they will probably still exist.

This type of mechanism did not lead to most of the complications
found in this paper, however. These were instead due to couplings
between the charge and 3-current densities as viewed in the
center-of-mass frame. In gravity, the situation could be even worse.
Similar interactions might exist between mass, 3-momentum, and
stress densities. Intuitively, though, it would appear that adopting
appropriate energy conditions should considerably soften these
interactions.

\begin{appendix}

\section{Dixon's Formalism}
\label{Dixon}

\subsection{Current Multipoles}
\label{CurrentMoments}

This portion of the appendix reviews Dixon's decomposition of the
electromagnetic current vector into multipole moments. It also
derives natural `potential functions' that can be used to generate
sets of moments for all physically interesting current vectors
satisfying (\ref{chargecons}). These objects are shown to determine
$J^{a}$ in a simple way. The main reasons for these constructions
are explained in Sec. \ref{discussion}, although some secondary
points are also mentioned in Sec. \ref{LawsMot}. In short, analogs
of these steps would become essential in the gravitational
self-force problem, so we include them here to allow a relatively
straightforward generalization.

The notation here will be that defined in Sec. \ref{LawsMot}. Using
it, we can define multipole moments for the current vector. Such
objects are usually constructed by integrating a source function
against a suitable number of radius vectors. Defining
$r^{a}=x^{a}-z^{a}(s)$ for $x \in \Sigma(s)$, one might therefore
expect that for $n \geq 0$, the $2n$-pole moment could be given by
\begin{equation}
Q^{b_{1} \cdots b_{n} a}(s) := \int \! \mathrm{d}^{4}x \, r^{b_{1}}
\cdots r^{b_{n}} \hat{J}^{a}\big(x-z(s),s\big) ~, \label{QJ}
\end{equation}
where we have assumed the existence of some distribution
$\hat{J}^{a}(x,s)$ which is related to $J^{a}(x)$, but has compact
support in $x$. This will be called the current skeleton. Note that
(\ref{QJ}) automatically implies that
\begin{equation}
Q^{b_{1} \cdots b_{n} a}=Q^{(b_{1} \cdots b_{n}) a}
\label{chargesym2}
\end{equation}
for all $n \geq 1$.

It is rather cumbersome to keep track of each $Q^{\cdots}$ directly,
so we instead define a generating function
\begin{equation}
G^{a}(k,s) := Q^{a}(s) + \sum_{n=1}^{\infty} \frac{(-i)^{n}}{n!}
k_{b_{1}} \cdots k_{b_{n}} Q^{b_{1} \cdots b_{n} a}(s) ~.
\label{chargegenerate}
\end{equation}
An arbitrary multipole moment can now be extracted from this in the
usual way:
\begin{equation}
Q^{b_{1} \cdots b_{n}  a}(s) =i^{n} \left. \Big( \partial^{b_{1}}
\cdots \partial^{b_{n}} G^{a}(k,s) \Big) \right|_{k=0} ~.
\label{momentderive}
\end{equation}
$G^{a}$ is therefore completely equivalent to the set $\{ Q^{a} ,
Q^{ba} , \ldots \}$.

Although this is a useful property, the definition of $G^{a}$ is not
simply a mathematical convenience. Using (\ref{QJ}) and
(\ref{chargegenerate}), it can be shown that
\begin{equation}
G^{a}(k,s) = \int \! \mathrm{d}^{4} r \, \hat{J}^{a}(r,s) e^{-i k
\cdot r} ~.
\end{equation}
The Fourier transform of $G^{a}$,
\begin{equation}
\widetilde{G}^{a}(r,s) := \int \! \mathrm{d}^{4}k \, G^{a}(k,s) e^{i
k \cdot r} ~,
\end{equation}
is therefore proportional to $\hat{J}^{a}$:
\begin{equation}
\widetilde{G}^{a}(r,s) = (2 \pi)^{4} \hat{J}^{a}\left(r,s\right) ~.
\label{Jhatdefine}
\end{equation}
This shows that if $J^{a}$ and $\hat{J}^{a}$ equivalent in an
appropriate sense, the set of moments can be used to completely
reconstruct the current vector.

In order to relate the current to its skeleton, it is convenient to
think of $\hat{J}^{a}$ as a linear functional on the space of all
$C^{\infty}$ test functions with compact support (as is typically
done in distribution theory). In particular, knowing
\begin{equation}
\left\langle \hat{J}^{a}(r,s), \phi_{a}(x) \right\rangle := \int \!
\mathrm{d}^{4} x \, \hat{J}^{a}(r,s) \phi_{a}(x)
\end{equation}
for all suitable test functions $\phi_{a}(x)$ can be used to define
$\hat{J}^{a}$. An analogous statement can also be made for $J^{a}$.
These two objects can then be related to each other by writing
$\langle J^{a},\phi_{a} \rangle$ in terms of $\langle \hat{J}^{a},
\phi_{a} \rangle$. The latter expression depends on $s$, while the
former one does not. It is therefore most straightforward to link
the two by simply integrating out the $s$-dependence:
\begin{eqnarray}
\Big\langle J^{a}(x), \phi_{a}(x) \Big\rangle &=& \int \!
\mathrm{d}s \, \left\langle \hat{J}^{a}(r,s), \phi_{a}(x)
\right\rangle ~, \label{JG1}
\\
&=& \frac{1}{(2\pi)^{4}} \int \! \mathrm{d}s \, \left\langle
G^{a}(k,s), \widetilde{\phi}_{a}(k) e^{-i k \cdot z(s)}
\right\rangle ~. \label{momentcurrent}
\end{eqnarray}
Following \cite{Dix70b}, we take this (along with
(\ref{chargegenerate})) to \textit{define} what is meant by saying
that that the $Q^{\cdots}$'s are `multipole moments of $J^{a}$.'

This is not unique definition, however. To remove the remaining
freedom in a useful way, we simply state Dixon's results
\cite{Dix67,Dix70b,Dix74}. Let
\begin{eqnarray}
n_{b_{1}} Q^{b_{1} \cdots b_{n-1} [b_{n} a]} &=& 0 ~,
\label{chargesym1}
\\
Q^{(b_{1} \cdots b_{n})} &=& 0 ~, \label{chargesym3}
\end{eqnarray}
for all $n \geq 2$. Also assume that the monopole moment has the
special form
\begin{equation}
Q^{a} = q v^{a} ~, \label{chargemono}
\end{equation}
where $q$ is the total charge as it is usually defined.

Now choose a test function of the form $\phi_{a}(x) = \partial_{a}
\phi(x)$, with $\phi(x)$ itself also a test function. Then
(\ref{momentderive}), (\ref{momentcurrent}), and (\ref{chargesym3})
can be used to show that
\begin{eqnarray}
\big\langle \partial_{a}J^{a} , \phi \big\rangle &=& -\big\langle
J^{a} , \partial_{a} \phi \big\rangle ~, \nonumber
\\
&=& \frac{i}{(2\pi)^{4}} \int \! \mathrm{d}s \, \left\langle k_{a}
G^{a}(k,s) , \tilde{\phi}(k) e^{-i k \cdot z(s)} \right\rangle ~,
\nonumber
\\
&=& - \int \! \mathrm{d}s \, q v^{a}(s)
\partial_{a}\phi\big(z(s)\big) ~, \nonumber
\\
&=& - \int \! \mathrm{d}s \, q \frac{\mathrm{d}}{\mathrm{d}s}
\phi\big(z(s)\big)~. \label{ChargeConsCheck}
\end{eqnarray}
This must vanish for all $\phi$, which can be ensured by simply
requiring that
\begin{equation}
\dot{q} = 0 ~. \label{qdot}
\end{equation}
This (trivial) evolution equation is the only one implied by
(\ref{chargecons}). It can be shown that moments satisfying
(\ref{chargesym2}), (\ref{chargesym1})-(\ref{chargemono}), and
(\ref{qdot}) describe any $J^{a}$ with the given properties in a
uniquely simple way \cite{Dix67,Dix70b,Dix74}. A precise statement
of the theorem that was proven is contained in \cite{Dix74}.

These conditions allow a great deal of freedom in choosing different
moments. The fact that there is only one evolution equation implied
by (\ref{chargecons}) does not mean that the higher moments
necessarily remain constant. Rather, they can be given an almost
arbitrary time dependence. This is a reflection of the fact that we
have not yet chosen to model any particular type of matter.
Specifying how the higher moments change in time is essentially
equivalent to choosing an equation of state. Although this must be
given in order to have a well-defined initial value problem,
self-forces and self-torques can be written down without ever having
to explicitly evaluate the time derivatives of the current moments.
This allows us to derive equations of motion valid for a very large
class of systems.

In order to do so, we first need to pick out sets of moments (or
equivalently their generating functions) which represent physically
reasonable current vectors. Despite appearances, this cannot be done
arbitrarily. Dixon's theorem ensures that any nonsingular current
vector with support $W$ can be described by a set of moments with
the given properties, although the reverse is not necessarily true.
Extra conditions need to be imposed in order to ensure that the
$J^{a}$ associated with any particular $G^{a}$ (or $\hat{J}^{a}$)
has the correct smoothness and support properties.

To gain some insight into this, fix some test function $\phi_{a}$.
From this, construct a second test function $\phi'_{a}$ which agrees
with $\phi_{a}$ everywhere except in an infinitesimal neighborhood
of $Z$. Assume that the support of $\phi'_{a}$ does not include $Z$
itself. (\ref{chargegenerate}) and (\ref{momentcurrent}) then show
that any finite number (and only finite number) of multipoles can be
changed without affecting $\left\langle J^{a},\phi'_{a}
\right\rangle$. The same cannot be said for $\left\langle
J^{a},\phi_{a} \right\rangle$. But for any physical current vector,
$\left\langle J^{a}, \phi'_{a} \right\rangle \simeq \left\langle
J^{a},\phi_{a} \right\rangle$. This shows that any finite subset of
an admissible collection of moments is completely determined by its
complement.

Because of this, it is not reasonable to impose conditions directly
on the individual moments to ensure that their associated current
vector is physically acceptable. Such restrictions are most easily
stated in terms of $\widetilde{G}^{a}$ or $\hat{J}^{a}$, but doing
so first requires finding how $\hat{J}^{a}$ is affected by
(\ref{chargesym1})-(\ref{chargemono}), and (\ref{qdot}). This is now
done by finding the general form of $G^{a}$, and then taking its
Fourier transform.

It is shown in \cite{Dix74} that the constraint equations are
precisely equivalent to requiring that the generating function have
the form
\begin{equation}
G^{a}(k,s)= G_{(1)}^{a}(h \cdot k,s) + (n \cdot k) G_{(2)}^{a}(h
\cdot k,s) ~, \label{G1G2}
\end{equation}
where $n \cdot k := n_{a} k^{a}$, $(h \cdot k)^{a}=h^{a}_{b} k^{b}$,
and $n_{a} G^{a}_{(2)}=0$. Also,
\begin{eqnarray}
k_{a}G^{a}(k,s)&=&q k_{a}v^{a}(s) ~, \label{Gtimesk}
\\
\partial^{[a} G^{b]}_{(2)}(h \cdot k,s) &=& 0 ~.
\label{dG2}
\end{eqnarray}

It is now useful to take Fourier transforms of these equations to
find their equivalent forms when representing the moments by
$\hat{J}^{a}$ ($= \widetilde{G}^{a}/(2\pi)^{4}$).
$\widetilde{G}^{a}_{(1)}$ doesn't take on any special form, although
(\ref{dG2}) shows that
\begin{equation}
\left\langle r^{c} h_{c}^{[a} \widetilde{G}^{b]}_{(2)}(r,s), \phi(r)
\right\rangle =0 ~. \label{G2constrain}
\end{equation}
This equation is solved by any $\widetilde{G}^{a}_{(2)}$ of the form
\begin{equation}
\widetilde{G}^{a}_{(2)}(r,s) = h^{a}_{b}(s) r^{b}
\widetilde{G}_{(2)}(r,s) ~, \label{G2form}
\end{equation}
for all functions $\widetilde{G}_{(2)}$ (Despite the notation, it
will shortly be clear that the inverse Fourier transform of
$\widetilde{G}_{(2)}$ does not exist in general.). Note that this is
not the most general solution of (\ref{G2constrain}). A term of the
form $g^{a}(n \cdot r, s) \delta^{3} (h \cdot r)$ may also be added
to $\widetilde{G}^{a}_{(2)}$, although we choose to ignore this
possibility.

In any case, (\ref{G1G2}) and (\ref{G2form}) give an explicit form
for $\widetilde{G}^{a}$
\begin{eqnarray}
\widetilde{G}^{a}(k,s)  &=& \widetilde{G}^{a}_{(1)} -i
h^{a}_{b}r^{b} n^{c} \partial_{c} \widetilde{G}_{(2)} ~.
\label{G2fourier}
\end{eqnarray}

Since $G_{(1)}^{a}$ and $G_{(2)}^{a}$ are independent of $n \cdot
k$, their (four-dimensional) Fourier transforms must be proportional
to $\delta(n \cdot r)$. It is therefore natural to define quantities
$A$, $B^{a}$, and $C$ such that
\begin{eqnarray}
\widetilde{G}^{a}_{(1)}(r,s) &=& (2 \pi)^{4} \delta(n \cdot r) \Big(
A(h \cdot r,s) n^{a} + B^{a}(h \cdot r,s) \Big)  ~,
\label{rhojdefine}
\\
\widetilde{G}_{(2)}(r,s) &=& -i(2 \pi)^{4} \delta(n \cdot r) C(h
\cdot r,s) ~, \label{Adefine}
\end{eqnarray}
where $B^{a}n_{a}=0$. Combining these expressions with
(\ref{G2fourier}), $\hat{J}^{a}$ is found to have the form
\begin{equation}
\hat{J}^{a}(r,s) = \delta(n \cdot r) \Big( A(h \cdot r,s) n^{a}(s) +
B^{a}(h \cdot r,s) \Big) - \delta'(n \cdot r) C(h \cdot r,s)
h^{a}_{b}(s) r^{b} ~. \label{hatJ}
\end{equation}

This is further restricted by (\ref{Gtimesk}), the Fourier transform
of which becomes
\begin{eqnarray}
\left\langle \partial_{a} \hat{J}^{a}(x,s) , \phi(x) \right\rangle
&=& (2\pi)^{-4} i \left\langle k_{a} G^{a}(k,s) ,
\widetilde{\phi}(k) \right\rangle ~, \nonumber
\\
&=& - (2\pi)^{-4} q v^{a}(s) \left\langle \widetilde{1} ,
\partial_{a} \phi(r) \right\rangle ~, \nonumber
\\
&=& q v^{a}(s) \Big\langle \partial_{a} \delta^{4}(r) , \phi(r)
\Big\rangle ~. \label{divG}
\end{eqnarray}
Comparing this to the divergence of (\ref{hatJ}) shows that
\begin{eqnarray}
q \delta^{3}(h \cdot r) &=& A(h \cdot r,s) - h^{a}_{b} \partial_{a}
\Big( r^{b} C(h \cdot r,s) \Big) ~, \label{Gconstrain1}
\\
q v^{a} \partial_{a} \delta^{3}(h \cdot r) &=& \partial_{a}B^{a} (h
\cdot r,s) ~. \label{Gconstrain2}
\end{eqnarray}

The solution to the `homogeneous' analog of (\ref{Gconstrain1}),
\begin{equation}
h^{a}_{b} \partial_{a} \Big( r^{b} C_{(0)} (h \cdot r,s) \Big) = - q
\delta^{3}(h \cdot r) ~,
\end{equation}
is
\begin{equation}
C_{(0)}(h \cdot x,s)= - \frac{q}{4 \pi |r|^{3}}~, \label{Csing}
\end{equation}
where $|r|^{2}:=-h_{ab} r^{a} r^{b} \geq 0$. For future convenience,
we now define new functions $\varphi(r,s)$ and $q_{0}(s)$ such that,
\begin{equation}
C= N \left( \varphi - \frac{q_{0}(s)}{4\pi |r|^{3}} \right) ~.
\label{Cform}
\end{equation}
$N$ is just the lapse, as given in (\ref{lapse}).

In terms of $q_{0}$ and $\varphi$, (\ref{Gconstrain1}) now becomes
\begin{equation}
A(h \cdot r,s) = h^{a}_{b} \partial_{a} \Big( r^{b} N \varphi(h
\cdot r, s) \Big) +\Big(q-q_{0}(s)\Big) \delta^{3}(h \cdot r) +
\frac{q_{0}}{4\pi} \frac{\dot{n}^{a} r_{a}}{|r|^{3}} ~. \label{rhoA}
\end{equation}
$\hat{J}^{a}$ is therefore given by (\ref{hatJ}) with $A$ and $C$
having the respective forms (\ref{rhoA}) and (\ref{Cform}). We also
have that $B^{a}$ satisfies (\ref{Gconstrain2}). This effectively
parameterizes all sets of moments satisfying Dixon's constraints.

(\ref{JG1}) can now be used to relate $\hat{J}^{a}$ to $J^{a}$. It
is convenient to do this in the $(r^{\alpha},\tau)$ coordinates
defined in Sec. \ref{LawsMot}. Using them, (\ref{JG1}) becomes
\begin{equation}
\left\langle J^{a}, \phi_{a} \right\rangle = \int \! \mathrm{d}s \!
\int \! \mathrm{d}^{3}r \! \int \! \mathrm{d}\tau \, N(x)
\hat{J}^{a}\big(x-z(s),s\big) \phi_{a}(x) ~. \label{JG2}
\end{equation}
Applying (\ref{hatJ}), the $\tau$-integral in this equation can be
carried out explicitly:
\begin{equation}
\left\langle J^{a}, \phi_{a} \right\rangle = \int \! \mathrm{d}s \!
\int \! \mathrm{d}^{3}r \: \Bigg\{ \bigg[ A n^{a} + B^{a} + N^{-2}
e^{a}_{\alpha} r^{\alpha} \dot{n}_{\beta} v^{\beta} C + N^{-1}
e^{a}_{\alpha} v^{\beta} \partial_{\beta} \big( r^{\alpha} C \big)
\bigg] \phi_{a} + N^{-1} C e^{a}_{\alpha} r^{\alpha} \frac{\partial
\phi_{a}}{\partial s} \Bigg\} ~, \label{JG3}
\end{equation}
where all quantities are evaluated at $s$.

Commuting the $s$-integral with the spatial ones and integrating the
last term by parts,
\begin{eqnarray}
\left\langle J^{a}, \phi_{a} \right\rangle &=& \int \!
\mathrm{d}^{3}r \! \int \! \mathrm{d}s \: \Bigg\{ n^{a} \left[A +
\frac{C}{N} \dot{n}_{\beta} r^{\beta} \right] + e^{a}_{\alpha}
\left[ B^{\alpha}+ v^{\beta} \partial_{\beta} \left( r^{\alpha}
\frac{C}{N} \right) - r^{\alpha} \frac{\partial}{\partial s} \left(
\frac{C}{N} \right) \right] \Bigg\} \phi_{a} ~. \label{JG4}
\end{eqnarray}
But we also have that $\left\langle J^{a}, \phi_{a} \right\rangle =
\int \! \mathrm{d}^{3}r \! \int \! \mathrm{d}s \, N J^{a} \phi_{a}$,
which gives us an obvious way to explicitly write $J^{a}$ in terms
of $A$, $B^{a}$, and $C$.

The charge density with respect to the tetrad frame takes on the
particularly simple form
\begin{eqnarray}
\rho &:=& n_{a}J^{a} ~,
\\
&=& \partial_{\alpha}\big( r^{\alpha} \varphi \big) + N^{-1}
(q-q_{0}) \delta^{3}(r) ~. \label{chargedensity2}
\end{eqnarray}
Physically, $\rho(r^{\alpha},s)$ must be nonsingular and have
support $W$. Without loss of generality, we can therefore choose
$q_{0}=q$. $\rho$ will then be admissible if $\varphi(r^{\alpha},s)$
is a continuous function.

$\hat{J}^{a}$ was originally introduced as something with compact
support (in $r^{a}$), so $A$, $B^{a}$, and $C$ must also have
compact support (in $r^{\alpha}$). Using this along with the
requirement that $\rho$ vanish outside $W$ implies that $\varphi =
q/4\pi |r|^{3}$ in this region. So any choice of $\varphi$ which
satisfies this `boundary condition' and is continuous will generate
a physically reasonable charge density.

For the 3-current $j^{\alpha} := e^{\alpha}_{a} J^{a}$, the
(necessarily regular) portions contributed by $\varphi$ can be
temporarily ignored to show that
\begin{equation}
j^{\alpha} = N^{-1} \left[ B^{\alpha} + \big(\ldots\big) - v^{\beta}
\partial_{\beta} \left( \frac{q r^{\alpha}}{4\pi |r|^{3}} \right)
\right] ~.
\end{equation}
The divergence of this last term is equal to $-q v^{\alpha}
\partial_{\alpha} \delta^{3}(r)$ , which exactly cancels the
divergence of $B^{\alpha}$ given in (\ref{Gconstrain2}). It is
therefore useful to write $B^{\alpha}$ in the form
\begin{equation}
B^{\alpha} = H^{\alpha} + v^{\beta} \partial_{\beta} \left( \frac{q
r^{\alpha}}{ 4\pi |r|^{3}} \right) ~, \label{Bform}
\end{equation}
with $H^{\alpha}$ an arbitrary piecewise continuous vector field
satisfying $\partial_{\alpha} H^{\alpha} =0$. This ensures that
$j^{\alpha}$ is nonsingular.

Writing out the current in full,
\begin{equation}
j^{\alpha} = N^{-1} \Big( H^{\alpha} + v^{\beta} \partial_{\beta}
\big( r^{\alpha} \varphi \big) - r^{\alpha} \dot{\varphi} \Big) ~,
\label{threecurrent2}
\end{equation}
it is clear that $j^{\alpha}$ will have support $W$ if
\begin{equation}
H^{\alpha}= -v^{\beta} \partial_{\beta} \left( \frac{q r^{\alpha}
}{4 \pi |r|^{3} } \right) \label{HBoundary}
\end{equation}
outside $W$. This also guarantees that supp$(B^{\alpha})=W$, as
required.

This completes our study of the current moments. Essentially all
physically interesting expansions can be extracted from a
$\hat{J}^{a}$ of the form (\ref{hatJ}). $A$, $B^{a}$ and $C$ are all
functions of $(r^{\alpha},s)$, and have support $W$. $C$ has the
form (\ref{Cform}), where $\varphi$ is an arbitrary continuous
function equalling $q/4\pi |r|^{3}$ outside of $W$. (\ref{rhoA})
shows that $A$ is also derived from $\varphi$  (with $q_{0}=q$).
$B^{a}$ has the form (\ref{Bform}), where $H^{\alpha}$ is a
piecewise continuous (3-) vector field satisfying $\partial_{\alpha}
H^{\alpha} =0$. The relevant portions of these results are also
summarized in Sec. \ref{LawsMot}.

\subsection{Stress-Energy Moments}

Let the multipole moments of $T^{ab}$ be denoted by the set $\{
t^{bc}, t^{a b c}, \ldots, t^{a_{1} \cdots a_{n} bc}, \ldots \}$,
where $t^{(a_{1} \cdots a_{n}) bc} = t^{a_{1} \cdots a_{n} (bc) } =
t^{a_{1} \cdots a_{n} bc}$. As with the current moments, it is
convenient to keep track of this collection with a generating
function
\begin{equation}
G^{ab}(k,s) := \sum_{n=0}^{\infty} \frac{(-i)^{n}}{n!} k_{c_{1}}
\cdots k_{c_{n}} t^{c_{1} \cdots c_{n} ab}(s) ~.
\label{stressgenerate}
\end{equation}

(\ref{stresscons}) inextricably links $T^{ab}$ and $J^{a}$, so an
expression for $\langle T^{ab}, \phi_{ab} \rangle$ as simple as the
one for $\langle J^{a}, \phi_{a} \rangle$ is not possible while
retaining simple constraint and evolution equations. Still, one
might expect that $\langle T^{ab}, \phi_{ab} \rangle$ should at
least be proportional to $\int \! \mathrm{d}s \, \langle
\hat{T}^{ab},\phi_{ab} \rangle$ (where $\hat{T}^{ab} := (2\pi)^{4}
\widetilde{G}^{ab}$). Define a distribution
$\widetilde{\Phi}^{ab}=\widetilde{\Phi}^{(ab)}$ to make up the
difference:
\begin{equation}
\left\langle T^{ab}, \phi_{ab} \right\rangle = \int \! \mathrm{d}s
\, \left\langle \hat{T}^{ab}(r,s) + \widetilde{\Phi}^{ab}(r,s),
\phi_{ab}(x) \right\rangle ~. \label{momentstress}
\end{equation}

As before, we call all sets $\{t^{\cdots}\}$ satisfying
(\ref{stressgenerate}) and (\ref{momentstress}) `multipole moments
of $T^{ab}$.' This is not a unique definition, however (even if
$\widetilde{\Phi}^{ab}$ were given). Following \cite{Dix67,Dix74},
we will now pick out a set which very simply and naturally implies
(\ref{stresscons}).

Using (\ref{ChargeConsCheck}) as a guide, $\left\langle
\partial_{a}T^{ab}, \phi_{b} \right\rangle$ can be found by
substituting a test function of the form $\phi_{ab} = \partial_{a}
\phi_{b}$ into (\ref{momentstress}). It is clear that the resulting
expression  depends on $k_{a}G^{ab}$, which is analogous to what
happened when computing $\partial_{a} J^{a}$. In that case,
(\ref{chargesym3}) showed that $k_{a} G^{a}$ depended only on
$Q^{a}$. (\ref{chargecons}) is usually interpreted as an expression
of global charge conservation, so it was not unreasonable that it
only restricted the monopole moment. In the case of the
stress-energy tensor, we expect that (\ref{stresscons}) should have
something to say about both the linear and angular momenta of the
body (i.e. its monopole and dipole moments). We therefore suppose
that $k_{a}G^{ab}$ involves only $t^{ab}$ and $t^{abc}$. This can be
accomplished by letting
\begin{equation}
t^{(a_{1} \cdots a_{n} b)c} = 0 \label{StressCon1}
\end{equation}
for all $n \geq 2$.

Using this constraint implies that
\begin{eqnarray}
\left\langle \partial_{b}T^{ab}, \phi_{a} \right\rangle & \propto &
\frac{1}{(2\pi)^{4}} \int \! \mathrm{d}s \, \bigg\langle i k_{b}
G^{ab}(k,s), \widetilde{\phi}_{a}(k) e^{-i k \cdot z(s)}
\bigg\rangle  ~,
\\
&=& \int \! \mathrm{d}s \, \bigg[ t^{ab} \partial_{a} \phi_{b} \big(
z(s) \big) + t^{abc} \partial_{a} \partial_{b} \phi_{c} \big( z(s)
\big) \bigg] ~.
\end{eqnarray}

Dixon found that the first two moments can be given the special
forms \cite{Dix67,Dix74}
\begin{eqnarray}
t^{ab} &=& p^{(a} v^{b)} ~,
\\
t^{abc} &=& S^{a(b} v^{c)} ~, \label{stressdipole}
\end{eqnarray}
where we call $p^{a}$ the linear momentum, and $S^{ab}= S^{[ab]}$
the angular momentum. Using these expressions,
\begin{equation}
\left\langle \partial_{b}T^{ab}, \phi_{a} \right\rangle \propto \int
\! \mathrm{d}s \, \left[ \dot{p}^{a} \phi_{a} \big(z(s)\big) +
\frac{1}{2} \left( \dot{S}^{ab} - 2 p^{[a} v^{b]} \right)
\partial_{a} \phi_{b} \big(z(s)\big) \right]  . \label{hatTevolve}
\end{equation}
In the absence of an electromagnetic field (or a current), the
left-hand side of this equation vanishes, and the proportionality
sign becomes an equality (since $\widetilde{\Phi}^{ab}$ vanishes in
this case). Varying $\phi_{a}$ then recovers the standard equations
of motion for a free particle:
\begin{eqnarray}
\dot{p}^{a} &=& 0 ~,
\\
\dot{S}^{ab} &=& 2 p^{[a} v^{b]} ~.
\end{eqnarray}

The situation is of course much more complicated when a field is
present. Writing out $\left\langle \partial_{a} T^{ab},\phi_{b}
\right\rangle$ in full, (\ref{stresscons}) implies that
\begin{equation}
\int \! \mathrm{d}s \: \Bigg\{ \left[ \dot{p}^{a} \phi_{a} +
\frac{1}{2} \left( \dot{S}^{ab} - 2 p^{[a} v^{b]} \right)
\partial_{a} \phi_{b} \right] + \bigg\langle i k_{b} \Phi^{ab} ,
\widetilde{\phi}_{a} e^{-i k \cdot z(s)}\bigg\rangle\Bigg\} = -
\left\langle F^{ab}J_{b}, \phi_{a} \right\rangle ~. \label{stress1}
\end{equation}
The right hand side of this equation can now be written in terms of
the current moments. Using (\ref{momentcurrent}), the Fourier
convolution theorem, and a Taylor series, it can be shown to be
\cite{Dix67}
\begin{eqnarray}
\left\langle F^{ab} J_{b}, \phi_{a} \right\rangle &=&
\frac{1}{(2\pi)^{4}} \int \! \mathrm{d}s \, \left\langle
\sum_{n=0}^{\infty} \frac{(-i)^{n}}{n!} k_{c_{1}} \cdots k_{c_{n}}
\Psi^{c_{1} \cdots c_{n} a}, \widetilde{\phi}_{a}(k) e^{-i k \cdot
z(s)} \right\rangle ~, \label{forcedense}
\\
\Psi^{c_{1} \cdots c_{n} a}(s) &:=& \frac{1}{(2\pi)^{4}}
\left\langle \sum_{p=0}^{\infty} \frac{(-i)^{p}}{p!} l_{d_{1}}
\cdots l_{d_{p}} Q^{d_{1} \cdots d_{p} c_{1} \cdots c_{n} b} ,
\widetilde{F}^a{}_{b}(l) e^{-i l \cdot z(s)} \right\rangle ~.
\label{Psidefine}
\end{eqnarray}
Here, $\widetilde{F}^{ab}$ does not quite represent the Fourier
transform of $F^{ab}$, which is not well-defined. It is instead
equal to the Fourier transform of some function $^{*}F^{ab}$ which
coincides with the field in some neighborhood of $\Sigma(s) \cap W$,
but has compact support. If $^{*}F^{ab}$ is just as smooth as
$F^{ab}$, its precise form is irrelevant \cite{Dix67}.

(\ref{forcedense}) now makes it natural to interpret the
$\Psi^{\cdots}$'s as multipole moments of the force density exerted
on the body. We might therefore expect the net force to be
proportional to $\Psi^{a}$, and the net torque to $\Psi^{[ab]}$.
This identification can be made if $\Phi^{ab}$ has the form
\cite{Dix67}
\begin{equation}
\Phi^{ab}(k,s) = \frac{1}{(2\pi)^{4}} \left\{ \Psi^{(ab)} - i k_{c}
\left[ \Psi^{c(ab)} - \frac{1}{2} \Psi^{abc} \right] +
\sum_{n=2}^{\infty} \frac{(-i)^{n}}{n! n} k_{c_{1}} \cdots k_{c_{n}}
\left[ 2 \Psi^{c_{1} \cdots c_{n} (ab)} - \frac{n+2}{n+1}
\Psi^{(c_{1} \cdots c_{n} ab)} \right] \right\} ~. \label{stress2}
\end{equation}
Combining this with (\ref{stress1}) shows that
\begin{equation}
\int \! \mathrm{d}s \, \left[ \left( \dot{p}^{a} + \Psi^{a} \right)
\phi_{a} + \frac{1}{2} \left( \dot{S}^{ab} - 2 p^{[a} v^{b]} + 2
\Psi^{[ab]} \right) \partial_{a} \phi_{b} \right] = 0 ~.
\end{equation}
Given that this must hold for all possible choices of $\phi_{a}$, it
follows that
\begin{eqnarray}
\dot{p}^{a} &=& - \Psi^{a} ~, \label{fdefine}
\\
\dot{S}^{ab} &=& 2 \left( p^{[a} v^{b]} -  \Psi^{[ab]} \right) ~.
\label{tdefine}
\end{eqnarray}
By construction, these are the only evolution equations implied by
(\ref{stresscons}). If we impose one more constraint equation:
\begin{equation}
n_{a_{1}} t^{a_{1} \cdots a_{n-2} [a_{n-1} [a_{n} b] c]} = 0
\label{StressCon2}
\end{equation}
for $n \geq 3$, the chosen moments are unique in an appropriate
sense. They are also sufficiently general to describe all physically
interesting stress-energy tensors \cite{Dix74}. Note that
$\Phi^{ab}$ only depends on $J^{a}$ and $F^{ab}$, and that the
constraint equations (\ref{StressCon1}) and (\ref{StressCon2}) are
independent of these quantities. Portions of the stress-energy
tensor which depend on the current have therefore been completely
isolated from those which are not.

Changes in the higher moments may once again be interpreted as
`equation of state' (this identification is actually more direct in
this case). Their evolution is not completely arbitrary, however.
Besides respecting the constraint equations, they must also be
chosen so that $T^{ab}$ remains physically reasonable. These extra
restrictions would be provided by analogs of (\ref{chargedensity2})
and (\ref{threecurrent2}). Although these will not be derived here,
they can probably be constructed in a similar way. The presence of
the field makes their derivation more complicated, but it should
still be possible to repeat all of the steps carried out with the
current moments.

It suffices to note that for our purposes, this procedure has
resulted in particularly natural definitions for the stress-energy
moments -- most importantly the linear and angular momenta. It can
be shown that the choices made here imply that these momenta are
given by (\ref{pdefine}) and (\ref{Sdefine}) \cite{Dix67}.
Interestingly, the net force and torque do not depend on $G^{ab}$ in
any necessary way. They are apparently as independent of the details
of the body's internal structure as possible.

Before moving on, we can gain some insight into the force moments
defined by (\ref{Psidefine}). Only the first two of these are
important here, and it is straightforward to show that
\begin{eqnarray}
\Psi^{a}&=& \left\langle \hat{J}_{b}(r,s), F^{ab}(x) \right\rangle
~, \label{forceJhat}
\\
\Psi^{[ab]}&=& \left\langle \hat{J}_{c}(r,s), r^{[a} F^{b]c}(x)
\right\rangle ~. \label{torqueJhat}
\end{eqnarray}
Applying (\ref{hatJ}), these expressions take the more explicit
forms
\begin{eqnarray}
\Psi^{a} &=& \int \! \mathrm{d}^{3} r \: \Bigg\{ N J_{b} F^{ab} +
\frac{\partial}{\partial s} \left[ \left( \varphi - \frac{q}{4\pi
|r|^{3}} \right) e^{\beta}_{b} r_{\beta} F^{ab} \right] \Bigg\} ~,
\label{forcedefine}
\\
\Psi^{[ab]} &=& \int \! \mathrm{d}^{3} r \: \Bigg\{ N J_{c}
r^{\alpha} e^{[a}_{\alpha} F^{b]c} + \left( \varphi - \frac{q}{4\pi
|r|^{3}} \right) e^{\gamma}_{c} r_{\gamma} v^{[a} F^{b]c} +
\frac{\partial}{\partial s} \left[ \left( \varphi - \frac{q}{4\pi
|r|^{3}} \right) e^{\gamma}_{c} r_{\gamma} r^{\alpha}
e^{[a}_{\alpha} F^{b]c} \right] \Bigg\} ~, \label{torquedefine}
\end{eqnarray}
which were derived by a different method in Sec. \ref{LawsMot}.

Although (\ref{forcedefine}) and (\ref{torquedefine}) are exact,
they are rather difficult to interpret. Their meaning is made
considerably more transparent if the field can be expanded in a
Taylor series inside $\Sigma(s) \cap W$. Then (\ref{forceJhat}) and
(\ref{torqueJhat}) together with (\ref{QJ}) show that
\begin{eqnarray}
\Psi^{a}(s) &\simeq& \sum_{\ell=0}^{L} \frac{1}{n!} \, Q^{c_{1}
\cdots c_{\ell}b} \Big( \partial_{c_{1}} \cdots
\partial_{c_{\ell}} F^a{}_{b} \Big)_{z(s)} ~, \label{forcemult}
\\
\Psi^{ab}(s) &\simeq& \sum_{\ell=0}^{L} \frac{1}{n!} \, Q^{d_{1}
\cdots d_{\ell} a c} \Big( \partial_{d_{1}} \cdots
\partial_{d_{\ell}} F^b{}_{c} \Big)_{z(s)} ~. \label{torquemult}
\end{eqnarray}
When $F^{ab}$ is approximately constant throughout the charge, we
recover the Lorentz force law, $\dot{p}^{a} = - \Psi^{a} \simeq - q
F^{ab} v_{b}$. Unfortunately, these series are not useful when the
field varies considerably over $\Sigma(s) \cap W$. And this is
exactly what the self-field does.

\end{appendix}

\begin{acknowledgments}
I would like to thank Pablo Laguna for reading over this manuscript,
and for many helpful discussions leading up to the results presented
here. I also thank Amos Ori for useful comments. This work was
supported in part by NSF grant PHY-0244788 and the Center for
Gravitational Wave Physics funded by the National Science Foundation
under Cooperative Agreement PHY-0114375.
\end{acknowledgments}

\end{document}